\documentclass[12pt]{article}

\usepackage{amsmath,amsthm, amsfonts, amssymb, amsxtra, amsopn}
\usepackage{pgfplots}
\usepgfplotslibrary{colorbrewer}
\pgfplotsset{compat = 1.15, 
			 cycle list/Set1-8} 
\usetikzlibrary{pgfplots.statistics, pgfplots.colorbrewer} 
\usepackage{pgfplotstable}
\usepackage{graphicx,grffile}
\usepackage{multirow}
\usepackage{booktabs}
\usepackage{listings}
\usepackage{cmap}
\usepackage{colortbl}
\usepackage{adjustbox}
\usepackage{epsfig}

\usepackage[tableposition=top,font=small,skip=5pt]{caption}
\usepackage{subcaption}
\usepackage{makecell} 
\usepackage[explicit]{titlesec}

\PassOptionsToPackage{hyphens}{url}

\usepackage{hyperref}
\hypersetup{colorlinks=true,linkcolor=black,citecolor=black,urlcolor=blue,filecolor=black}
\hypersetup{pdfpagemode=UseNone,pdfstartview=}

\definecolor{darkgreen}{rgb}{0.125,0.5,0.169}
\usepackage{enumitem}
\setlist[itemize]{noitemsep, topsep=0pt}

\advance\oddsidemargin by -0.35in
\advance\textwidth by 0.7in

\advance\topmargin by -0.45in
\advance\textheight by 0.9in

\long\def\symbolfootnotetext[#1]#2{\begingroup%
\def\thefootnote{\fnsymbol{footnote}}\footnotetext[#1]{#2}\endgroup}

\newcommand\dunderline[3][-1pt]{{%
  \sbox0{#3}%
  \ooalign{\copy0\cr\rule[\dimexpr#1-#2\relax]{\wd0}{#2}}}}
\def\uuu{\kern-1pt\dunderline{0.75pt}{\phantom{M}}}

\DeclareMathOperator{\tp}{TP}
\DeclareMathOperator{\fp}{FP}
\DeclareMathOperator{\fn}{FN}
\DeclareMathOperator{\precision}{precision}
\DeclareMathOperator{\recall}{recall}

\def\Agensla{\texttt{Agensla}}
\def\Androm{\texttt{Androm}}
\def\Convagent{\texttt{Convagent}}
\def\Crypt{\texttt{Crypt}}
\def\Crysan{\texttt{Crysan}}
\def\DCRat{\texttt{DCRat}}
\def\Injuke{\texttt{Injuke}}
\def\Makoob{\texttt{Makoob}}
\def\Mokes{\texttt{Mokes}}
\def\Noon{\texttt{Noon}}
\def\Remcos{\texttt{Remcos}}
\def\Seraph{\texttt{Seraph}}
\def\SnakeLogger{\texttt{SnakeLogger}}
\def\Stealerc{\texttt{Stealerc}}
\def\Strab{\texttt{Strab}}
\def\Taskun{\texttt{Taskun}}
\def\Zenpak{\texttt{Zenpak}}

\def\zz{\phantom{0}}

\title{A Comparison of Selected Image Transformation Techniques for Malware Classification} 

\author{Rishit Agrawal\footnotemark[1]\ \ \ 
Kunal Bhatnagar\footnotemark[1]\ \ \ 
Andrew Do\footnotemark[1]\ \ \ 
Ronnit Rana\footnotemark[1]\\
Mark Stamp\footnotemark[1]\,\,\footnotemark[2]}

\begin{document}

\symbolfootnotetext[1]{Department of Computer Science, San Jose State University}
\symbolfootnotetext[2]{mark.stamp$@$sjsu.edu}

\maketitle

\abstract
Recently, a considerable amount of malware research has 
focused on the use of powerful image-based machine learning techniques, 
which generally yield impressive results. However, before
image-based techniques can be applied to malware, the samples must be converted
to images, and there is no generally-accepted approach for doing so. 
The malware-to-image conversion strategies found in 
the literature often appear to be ad hoc, with little or no effort made to take into 
account properties of executable files. In this paper,
we experiment with eight distinct malware-to-image conversion techniques,
and for each, we test a variety of learning models. We find that 
several of these image conversion techniques perform similarly across a 
range of learning models, in spite of the image conversion processes being quite different. 
These results suggest that the effectiveness
of image-based malware classification techniques may depend more on
the inherent strengths of image analysis techniques, as opposed to the 
precise details of the image conversion strategy.

\section{Introduction}

In this paper, we explore image-based malware classification techniques.
When using such techniques, inherently one-dimensional executable files 
are converted into multi-dimensional images. Many different executable-to-image
conversion techniques have appeared in the literature, and to the best of the authors'
knowledge, little effort has been made to compare the effectiveness of these
different approaches. The ad hoc nature of this important aspect of image-based malware
analysis is somewhat surprising, given the extensive research that has been conducted in
this domain in recent years.

Specifically, we investigate eight distinct executable-to-image conversion techniques, and
for each we experiment with seven different classifiers. For four of these classifiers, we
consider two distinct methods of extracting features from the malware images, while
three of the classifiers are trained directly on the malware images. We compare
these image-based results to baseline experiments, where models are trained on 
byte histograms. Our results show the inherent strength of the use of image-based
features in the malware domain. We also find that several of the image conversion
techniques yield comparable results across different models. 

The remainder of the paper is organized as follows.
In Section~\ref{sect:related}, we discuss selected examples of related work.
Relevant background topics are introduced in Section~\ref{sect:background},
including details on the dataset used, image transformation techniques, and the learning 
models considered. Section~\ref{sect:experimentprep} covers our experimental design 
and we outline the experiments that we conduct. 
We present and analyze our results in Section~\ref{sect:experimentresults},
while in Section~\ref{sect:conclusion}, we summarize our main findings, 
and we discuss future research directions.

\section{Related Work}\label{sect:related}

Image-based analysis has recently become a mainstay of research in the malware field.
The literature in this area is vast---here we simply provide representative
examples of results that are most relevant to the research presented in this paper.

The work presented in~\cite{10.1145/2016904.2016908} appears to be the first attempt
to apply image-based techniques to the malware classification problem. They convert
binaries to greyscale using an ad hoc approach, based on the size of the malware
executable. They generate a malware image dataset, MalImg, and they
obtain high accuracies using image classifiers on this dataset.

It is worth noting that the MalImg dataset has become 
a standard dataset in this research domain. 
For example, in~\cite{Sravani}, GIST descriptors are used to
extract features from the MalImg samples and high classification
accuracy is obtained using various types of machine learning models. 
However, the MalImg dataset only includes images---not the original
malware executables---which somewhat limits its utility, as different image conversion techniques
cannot be considered. For this reason, the MalImg dataset is not 
suitable for the research problem that we consider in this paper.

Transfer learning involving pre-trained models is widely used in image analysis.
Not surprisingly, transfer learning has been shown to be effective
for image-based malware classification~\cite{Niket}. Generative Adversarial Networks (GANs) have also
been shown to be an effective tool for malware analysis~\cite{Huy}. Many standard image types 
can be generated from malware, ranging from the straightforward 
grayscale MalImg images~\cite{10.1145/2016904.2016908} 
to images based on QR and Aztec codes~\cite{Atharva}. 

In the paper~\cite{XIAO202049}, entropy-based features are extracted from malware samples
and these features are then encoded in images. The research in~\cite{Vu} goes even further,
with a wide range of malware features encoded in a color image format.
Several of the image types that we considered in the present paper were inspired by the
research in~\cite{Vu}.

A common approach in the image-based malware 
literature is to compare a variety of learning models
trained on a fixed set of malware-derived images. A representative example of this 
type of research is~\cite{Prat}, where no less than eight different learning models
are tested, but only one method of generating images is considered.
As far as the authors are aware, analyzing the effectiveness of a range of distinct 
malware-to-image conversion processes is a relatively neglected aspect of
research in the image-based malware analysis domain.

\section{Background}\label{sect:background}

In this section, we first discuss the malware dataset that forms the basis for our experiments.
Next, we introduce the learning models that we consider. We then discuss two feature extraction
techniques we use in our experiments. Finally, we outline the executable-to-image conversion
techniques that we employ in our experiments.

\subsection{Malware Dataset}

It is well established that malware developers generally create new malware
samples by modifying existing malware. Due to this process, 
malware can be classified into distinct families, based on common functions
and origins~\cite{Aycock}.

The malware data we use for this paper was derived from the extensive
RawMalTF dataset~\cite{2025rawmaltfrawmalwaredataset},
which was constructed from malware samples obtained from VirusShare, 
MalwareBazaar, and VXunderground. Each sample includes a family label,
and there are~65 distinct malware families. For our experiments, we eliminated all
families with less than~1000 samples, resulting in the
following~17 malware families.

\begin{description}
\item[\Agensla]\!--- A Trojan that scans system files and registry entries for stored 
passwords~\cite{kasperskyagensla}.
\item[\Androm]\!--- A modular downloader-backdoor with anti-VM features that pulls down 
additional payloads~\cite{kasperskyandrom}.
\item[\Convagent]\!--- A Win32 backdoor that intercepts keystrokes~\cite{kasperskyconvagent}.
\item[\Crypt]\!--- A HackTool that can hide a malicious user's presence on a system~\cite{kasperskycrypt}.
\item[\Crysan]\!--- A backdoor that drops modular stealer payloads onto the 
victim’s system~\cite{kasperskycrysan}.
\item[\DCRat]\!--- A RAT that is offered as a service, enabling file management, remote shell, 
and webcam access~\cite{kasperskydcrat}.
\item[\Injuke]\!--- A Trojan that injects ransomware payloads into processes to silently encrypt 
documents before displaying a ransom note~\cite{kasperskyinjuke}.
\item[\Makoob]\!--- A Trojan spyware that logs keystrokes and screenshots for remote 
retrieval~\cite{kasperskymakoob}.
\item[\Mokes]\!--- Similar to Makoob, with a modular architecture that enables
new features~\cite{kasperskymokes}.
\item[\Noon]\!--- A generic Trojan spyware that can capture user activity, browser cookies, 
and system information~\cite{kasperskynoon}.
\item[\Remcos]\!--- A backdoor that enables silent surveillance and remote control of 
Windows hosts~\cite{kasperskyremcos}.
\item[\Seraph]\!--- A Trojan downloader known for targeting browser and application 
credentials~\cite{kasperskyseraph}.
\item[\SnakeLogger]\!--- A modular Trojan keylogger that stealthily captures user input and 
session cookies~\cite{kasperskysnakelogger}.
\item[\Stealerc]\!--- An infostealer malware, specializing in credential and crypto-wallet 
theft~\cite{kasperskystealerc}.
\item[\Strab]\!--- A generic Win32 Trojan capable of executing arbitrary commands on compromised 
machines~\cite{kasperskystrab}.
\item[\Taskun]\!--- A Trojan downloader that leverages Windows Task Scheduler to achieve stealthy 
persistence~\cite{kasperskytaskun}.
\item[\Zenpak]\!--- A Trojan backdoor capable of key logging, among other 
features~\cite{kasperskyzenpak}.
\end{description}

From the source dataset, we randomly selected a subset of~1000 samples from each of
these~17 malware families. In Table~\ref{tab:dataset}, 
we provide basic statistics on these~17,000 samples. Note that we use the
first~$224\times 224=50,176$ bytes of each sample to construct images,
padding with~0 bytes, if necessary.

\begin{table}[!htb]
\centering
\caption{Dataset bytes (1000 samples per family)} \label{tab:dataset}
\adjustbox{scale=0.825}{
\begin{tabular}{l|rrr|cc}
\toprule
\multirow{2}{*}{\textbf{Family}} & \multicolumn{3}{c|}{\textbf{Byte statistics}} 
	& \multicolumn{2}{c}{\textbf{Percentage}} \\
	& \multicolumn{1}{c}{\texttt{min}}
	& \multicolumn{1}{c}{\texttt{max}}
	& \multicolumn{1}{c|}{\texttt{mean}}
	& \multicolumn{1}{c}{\textbf{truncated}} 
	& \multicolumn{1}{c}{\textbf{padded}} \\
\midrule
\Agensla & 6144 & 91226112 & 1093099.62  & \zz98.8 & \zz1.2 \\
\Androm & 5632 & 57700088 & 886886.96 & \zz96.2 & \zz3.8 \\
\Convagent & 3584 & 76124537 & 5502830.06 & \zz99.1 & \zz0.9 \\
\Crypt & 5632 & 64182272 & 1430292.00 & \zz95.3 & \zz4.6 \\
\Crysan & 8192 & 79691776 & 896145.35 & \zz56.7 & 43.3 \\
\DCRat & 62976 & 23420388 & 1876588.70 & 100.0 & \zz0.0 \\
\Injuke & 5632 & 41295366 & 2931494.75 & \zz95.1 & \zz4.9 \\
\Makoob & 69632 & 8117278 & 699149.01 & 100.0 & \zz0.0 \\
\Mokes & 31046 & 2900256 & 338075.11 & \zz95.1 & \zz4.9 \\
\Noon & 6656 & 38000000 & 925278.94 & \zz98.5 & \zz1.4 \\
\Remcos & 6656 & 53477376 & 1051645.45 & \zz98.8 & \zz1.2 \\
\Seraph & 5632 & 104199168 & 1452149.65 & \zz86.3 & 13.7 \\
\SnakeLogger & 6144 & 9854976 & 845841.59 & \zz99.1 & \zz0.9 \\
\Stealerc & 6144 & 69376619 & 1381113.08 & \zz99.8 & \zz0.2 \\
\Strab & 10752 & 80740352 & 863362.09 & \zz98.2 & \zz1.7 \\
\Taskun & 380416 & 80740352 & 926169.97 & 100.0 & \zz0.0 \\
\Zenpak & 5632 & 14228480 & 748459.78 & \zz99.5 & \zz0.5 \\
\bottomrule
\end{tabular}%
}
\end{table}

From Table~\ref{tab:dataset} we observe that for all families except \Crysan\ and \Seraph, 
the vast majority of samples are truncated. Overall, the similarity of the statistics between families
indicates that this dataset will likely provide a challenging test case for our image 
conversion experiments.

\subsection{Learning Models}\label{sect:models}

We consider seven distinct learning models, including classic learning techniques,
deep learning models, and advanced pre-trained image-based models. In this
section, we provide a brief introduction to each of these models.

\subsubsection{K-Nearest Neighbors}

K-Nearest Neighbors (KNN) classifies samples based on the~$k$
``nearest'' samples in the training dataset~\cite{Cunningham_2021}. 
No explicit training phase is required in KNN. However, KNN can become 
computationally expensive during inference, especially on large datasets.
In addition, for small values of~$k$, the KNN technique tends to overfit,
and for this reason, it can be challenging to determine an optimal value for~$k$.

\subsubsection{Multi-Layer Perceptron}

Multi-Layer Perceptron (MLP) is a classifier built on a standard neural network.
MLPs can model complex non-linear relationships between inputs and outputs. 
Generically, an MLP consists of an input layer, one or more hidden layers 
involving non-linear activation functions, and an output layer. An MLP is trained 
iteratively using the backpropagation algorithm~\cite{MultilayerperceptronPopescu}.

\subsubsection{Support Vector Machine}

Support Vector Machine (SVM) represents a class of classifiers that attempt to find an optimal 
separating hyperplane between classes. SVMs can efficiently deal with non-linear decision 
boundaries, thanks to the so-called kernel trick~\cite{CERVANTES2020189}.

\subsubsection{Extreme Gradient Boosting}

Boosting is a general learning strategy, whereby relatively weak classifiers can be 
combined to yield a stronger classifier---under optimal conditions, an arbitrarily
strong classifier can be obtained.
Extreme Gradient Boosting (XGBoost) is a boosting technique 
that mitigates some of the inherent instability that affects
simpler boosting strategies, such as AdaBoost.
The XGBoost classifier includes algorithms for split finding, caching, and parallelism,
and the technique has yielded state-of-the-art results in many cases~\cite{Chen_2016}.

\subsubsection{Transfer Learning Models}

MLP, as described above, is an example of an Artificial Neural Network (ANN). 
Convolutional Neural Networks (CNN) are another form of ANN that are optimized
for image-based tasks~\cite{cnnintro}. 

Transfer learning consists of training a model on a large dataset, then fine-tuning the 
model (i.e., retraining only the output layer) for a specific task. Transfer learning has proven
extremely effective when used with advanced CNN architectures. In this paper, we
consider three CNN-based transfer learning models, which we now introduce.

Visual Geometry Group 16 (VGG16) is a popular and highly effective
computer vision model~\cite{VGG16}. 
VGG16 was designed as a deep convolutional neural network, and it had been pre-trained 
for image classification on the well-known ImageNet~\cite{imagenet} dataset.
%As the name suggests, the model includes~16 layers with trainable parameters. 
VGG16 has~13 convolutional layers, five max-pooling layers, and three dense layers, 
for a total of~21 layers. Of these~21 layers, the five 
max-pooling layers do not contain any trainable weights, and hence there 
are~16 layers with trainable parameters, whence the~``16'' in VGG16.

%One unique aspect of VGG16 is its architectural uniformity. It employs convolutional layers with a 
%consistent~$3\times 3$ filter size and a stride of one, using the same padding throughout. 
%Additionally, max-pooling layers in VGG16 use a~$2\times 2$ filter with a stride of two. 
%This simplicity facilitates ease of implementation and efficient training.

%The generalization ability of VGG16 to images beyond its training data has made it a popular
%and successful model. VGG16 is commonly employed in transfer learning, where the original 
%dense layers are replaced with new task-specific dense layers. The hidden layers, 
%consisting of the convolutional and max-pooling layers from the original model, 
%remain unchanged and are used as a feature extractor while training the new fully 
%connected layers on the new data.

%\subsubsection{Overview of DenseNet121}%%%%% Kelvin Jou

DenseNet121 is a specific convolutional neural network architecture from
the DenseNet family of models~\cite{dense}. This model
includes four dense blocks and several transition 
layers consisting of a mix of convolutional and pooling layers. 
An unusual feature of DenseNet models is that the dense layers receive direct input 
from all preceding layers within the same block, which enables feature reuse. 
Transition layers are inserted between dense blocks to control spatial dimensions 
and channel depth of feature maps. A dense block is typically followed by a pooling 
layer, which reduce the dimensionality. DenseNet121 ends with a 
fully connected layer that uses a softmax activation function.

%DenseNet121 was designed to address the limitations of traditional CNN architectures, 
%such as vanishing gradients and information flow constraints. Since its introduction in~2017, 
%the model has been successfully applied to image classification tasks and object detection. 
%With excellent information flow and feature reuse, DenseNet121 can capture fine-grained 
%details and small-scale patterns throughout the network, which is crucial for image analysis. 
%In spite of having more than six million trainable parameters,
%DenseNet121 is more computationally efficient and requires less memory
%than many other comparable CNN models, 
%including ResNet152 and VGG16~\cite{dense}.

%\subsubsection{Overview of InceptionV3}%%%%% Nathan Zhang

InceptionV3 is another popular and well-known CNN architecture 
that has been successful applied 
to problems in computer vision. This architecture was developed as an 
enhancement to Google's Inception model~\cite{inception}.
A unique feature of InceptionV3 is its proprietary ``Inception Modules.'' 
These Inception Modules incorporate convolution operations with distinct kernel sizes that operate 
simultaneously, thereby enabling the model to more efficiently learn features of the input data.

%In typical applications, the input to an InceptionV3 model comprises image data, and its output layer 
%delivers predictions across a pre-defined set of classes. The intervening layers of the 
%architecture---including numerous convolutional layers, pooling layers, Inception modules, and fully 
%connected layers---perform sequential transformations of the input data. This sequence facilitates 
%the extraction of patterns and relevant features from the images.

%The training of the InceptionV3 model employs backpropagation and gradient descent. Due 
%to its complex and deep structure, it also employs advanced techniques such as 
%batch normalization (BatchNorm) and sophisticated initialization schemes. 
%These approaches are intended to ensure efficient training and 
%mitigate potential issues such as the vanishing gradient problem.

%The InceptionV3 architecture is known for its balance of computational efficiency and high 
%accuracy, performing effectively even with a large number of classes and when 
%handling high-resolution image data. 
%Nevertheless, training the InceptionV3 network can be computationally 
%intensive, and typically requires a substantial volume of labeled data.

\subsection{Feature Extraction}\label{subsect:featureextraction}

For our CNN-based models, namely, VGG16, DenseNet121, InceptionV3, 
no feature extraction is required, as these models are trained directly on images.
In contrast, KNN, SVM, MLP, and XGBoost are trained on feature vectors.

We use KNN, SVM, MLP, and XGBoost for baseline experiments, where
these models are trained on simple features, namely,
normalized byte histograms, using all bytes in each sample. 
We also extract features from the generated malware images
and train each of these models on the extracted features. 
Specifically, we experiment with the following two techniques
to extract features from malware images.

\subsubsection{Histogram of Oriented Gradients}

Histogram of Oriented Gradients (HOG) is a method that divides an 
image into a gradient map, and then computes a histogram of the resulting oriented gradients. 
HOG is particularly useful for capturing edge and gradient information~\cite{BHATTARAI2023102747}.

\subsubsection{Haralick Texture Features}

As the name suggests, Haralick Texture Features deal with the texture of an image, 
and this feature is useful for distinguishing between surfaces. 
Since malware images visually differ in texture, 
this method may provide useful features~\cite{Boland_1999,4309314}.

\subsection{Image Transformation Techniques}\label{subsect:imagetransformations}

In this section, we introduce the eight malware-to-image transformation 
techniques that we consider in this paper. 
All of the techniques discussed in the section ultimately
convert a sequence of byte values into a~$224\times 224$ image.
%with the exception of the Spiral images presented in
%Section~\ref{subsect:spiral}, below.

\subsubsection{Grayscale}\label{subsect:grayscale}

\begin{figure}[!htb]
	\centering
        \includegraphics[width=0.475\textwidth]{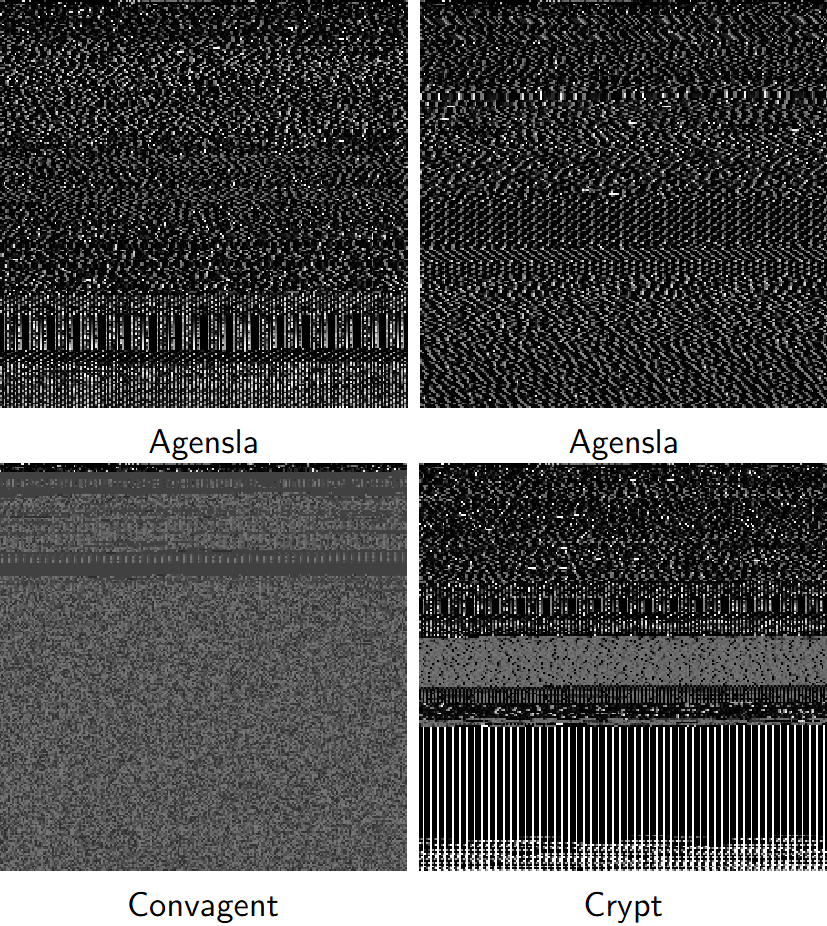}
        \caption{Grayscale images of Agensla, Convagent, and Crypt}
        \label{fig:viz_grayscale}
\end{figure}

Grayscale is the simplest approach to image conversion, and is often used in practice.
To create a grayscale 
image, we simply interpret the one-dimensional input array of~50,176 byte 
as a~$224\times 224$ two-dimensional array, truncating or padding with~0, as
necessary. Each value in this two-dimensional array represent the luminosity of an image 
pixel. Examples of Grayscale images from our dataset are given in 
Figure~\ref{fig:viz_grayscale}.

\subsubsection{Byteclass}\label{subsect:byteclass}

Our Byteclass method is somewhat analogous to Grayscale, except that 
we map integer values to green colors of 
varying luminosity. Thus, Byteclass can reveal insights into the structure of a binary 
file, and reveal details related to the distribution of various character classes~\cite{Vu}.
The specific encoding process is given in Table~\ref{tab:byteC}.

%%%%% Still not sure about this
\begin{table}[!htb]
\caption{Byteclass encodings}\label{tab:byteC}
\centering
\adjustbox{scale=0.85}{
\begin{tabular}{l|ll}
\toprule
\textbf{Bytes} & \textbf{Description} & \textbf{(R,G,B) encoding} \\ \midrule
0 & NULL & $(0, 0, 0)$ \\
1 -- 31 and 127 & ASCII control characters & $(0, 255, 0)$ \\
32 -- 126 and 128 -- 254 & printable ASCII characters & $(0, 32, 0)$ \\
%128 -- 254 & extended ASCII characters & $(0, 32, 0)$ \\
255 & extended ASCII character & $(0, 128, 0)$ \\
\bottomrule
\end{tabular}
}
\end{table}

\begin{figure}[!htb]
	\centering
        \includegraphics[width=0.475\textwidth]{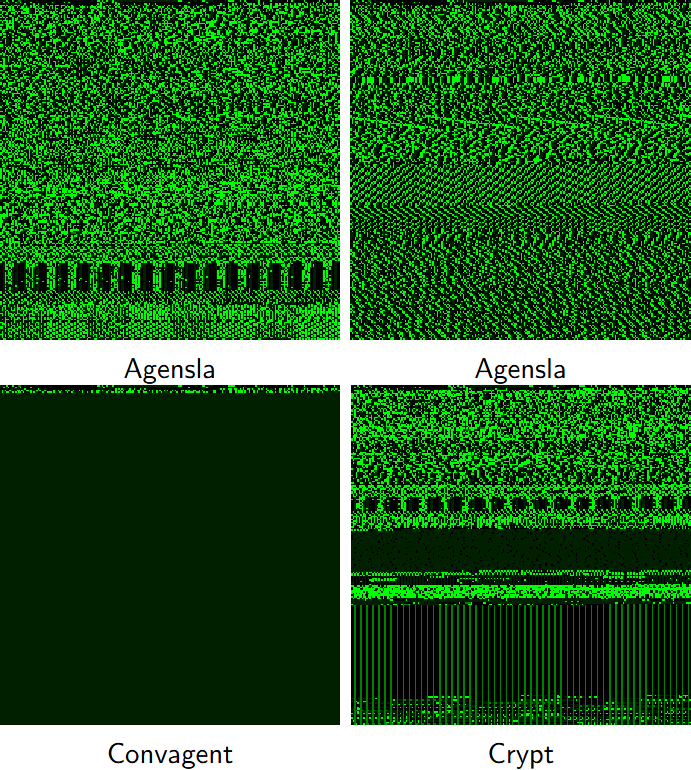}
        \caption{Byteclass images of Agensla, Convagent, and Crypt}
        \label{fig:viz_byteclass}
\end{figure}

Examples of Byteclass images from our dataset appear in Figure~\ref{fig:viz_byteclass}.
From Table~\ref{tab:byteC} note that we set the~R and~B bytes to~0, so that these images are 
based only on the~G component.

\subsubsection{Hilbert Curve}\label{subsect:hcurve}

A Hilbert curve is a layout that generates a space filling curve~\cite{keller2022renderinghilbertcurve}.
To generate a Grayscale image, when we reach the end of a line,
we simply continue at the start of next line, one row down.
In contrast, a Hilbert curve should better preserve locality information.

\begin{figure}[!htb]
\centering
\begin{tabular}{ccc}
  \includegraphics[width=0.3\linewidth]{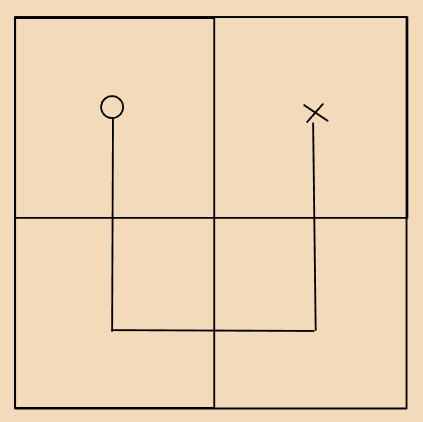}
  & &
  \includegraphics[width=0.3\linewidth]{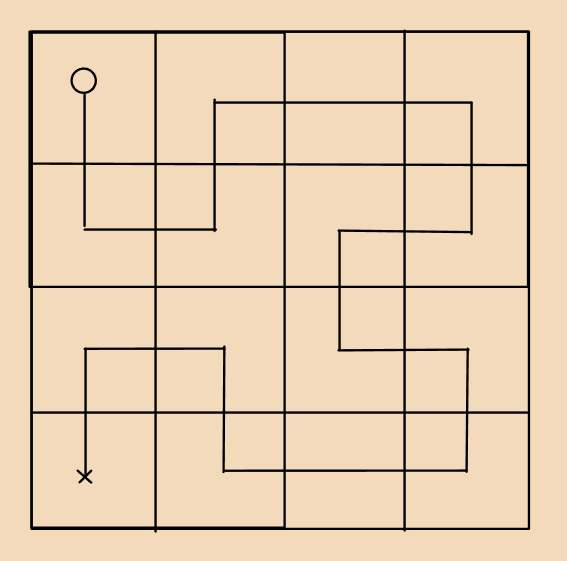}
  \\
  (a) 1st order hcurve
  & &
  (b) 2nd order hcurve
  \end{tabular}
  \caption{Hilbert curve example}\label{fig:hcurve}
\end{figure}

\begin{figure}[!htb]
	\centering
	\includegraphics[width=0.475\textwidth]{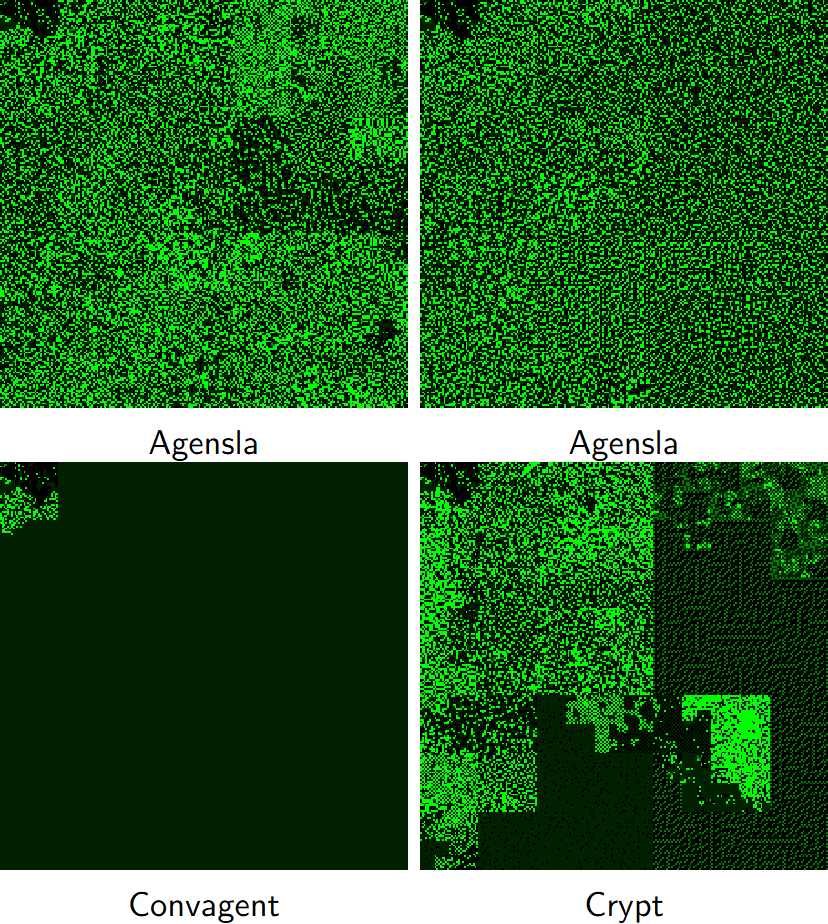}
	\caption{Hilbert curve images of Agensla, Convagent, and Crypt}
	\label{fig:viz_byteclass_hcurve}
\end{figure}

To generate a Hilbert curve, we start with a~$2\times 2$ square, which
is a first order Hilbert curve, as illustrated in Figure \ref{fig:hcurve}(a). 
To generate a second order Hilbert curve, each of the four quadrants in the first-order curve
is divided into four quadrants, yielding a~$4\times 4$ array,
with the data following the pattern illustrated in Figure \ref{fig:hcurve}(b).
This process is then repeated until we reach a size that will contain
all of the data values under consideration. In our case, we have~50,176
byte values to be put into a~$224\times 224$, and we use the Python library 
\texttt{hilbertcurve}~\cite{hilbertcurve} to generate our Hilbert images.
Examples of Hilbert images from our dataset appear in 
Figure~\ref{fig:viz_byteclass_hcurve}.

\subsubsection{Entropy}\label{subsect:entropy}

Our Entropy image conversion method provides a visual display of the 
entropy or uncertainty of the bytes~\cite{XIAO202049}.
Let~$b_i$, for~$i=0,1,\ldots,N$, be the bytes of a given sample. Then
the entropy value at position~$i$ is determined by computing the Shannon entropy of
bytes~$B=(b_i,b_{i+1},\ldots,b_{i+n})$, where~$n=\min\{255,N-i\}$. Shannon entropy is
computed as
$$
  H=-\sum_{i=0}^{255} P_i\,\log_2\,P_i
$$
where~$P_i$ is the relative frequency of byte value~$i$ in the block~$B$.

The process outlined in the previous paragraph
yields an array of entropy values~$x_0,x_1,\ldots,x_{255}$ 
which are then encode it into an image in the~$R=\{r_i\}$ and~$B=\{b_i\}$
planes of an RGB image, with the encoding given by~\cite{Vu}
$$
  r_i = 
    \left\{
    \begin{array}{ll}
        \Big\lfloor 256\big((x_i - \frac{1}{2}) - (x_i - \frac{1}{2})^2\big)^4\Big\rfloor 
        			& \mbox{if } x_i > \frac{1}{2}\\[1.25ex]
        0 & \text{otherwise}
    \end{array}
    \right.
$$
and
$$ 
  b_i = \big\lfloor 255\cdot x_i^2\big\rfloor .
$$

\begin{figure}[!htb]
	\centering
        \includegraphics[width=0.475\textwidth]{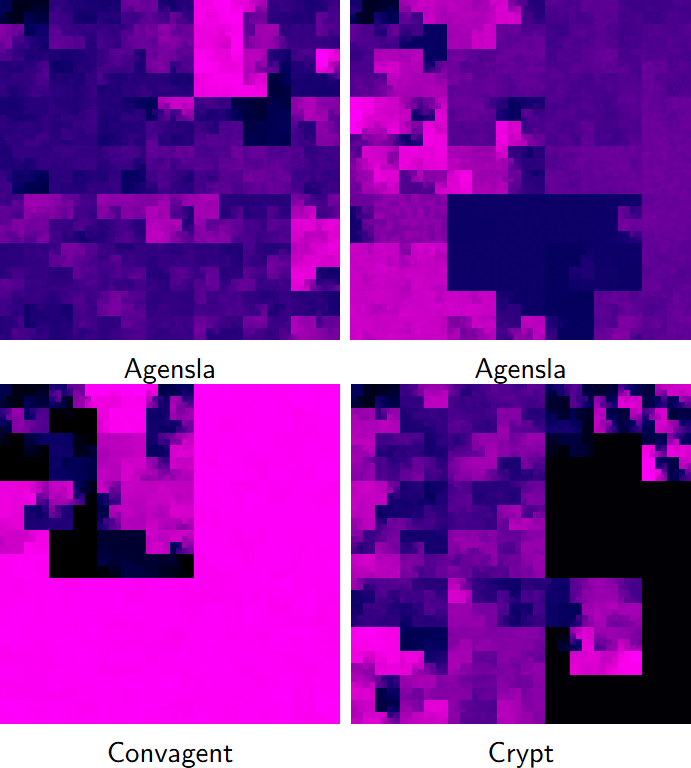}
        \caption{Entropy images of Agensla, Convagent, and Crypt}
        \label{fig:viz_entropy}
\end{figure}

To convert this sequence into a two-dimensional image, we rasterize the values 
onto a~$224\times 224$ canvas using a Hilbert curve traversal 
(as discussed in Section~\ref{subsect:hcurve}), which preserves local byte adjacency.
Examples of entropy images from our dataset appear in Figure~\ref{fig:viz_entropy}.

\subsubsection{HIT}\label{subsect:HIT}

\begin{figure}[!htb]
	\centering
	\includegraphics[width=0.475\textwidth]{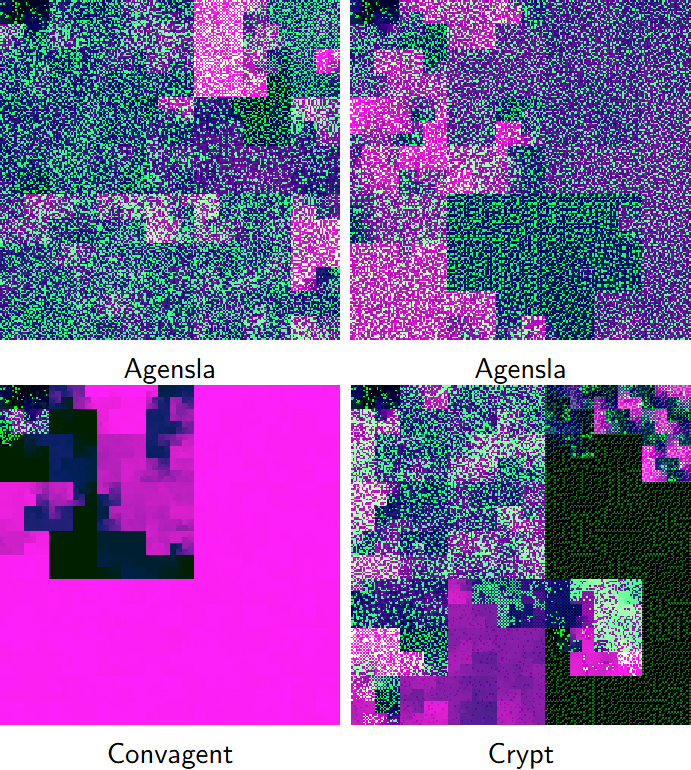}
	\caption{HIT images of Agensla, Convagent, and Crypt}
        \label{fig:viz_hit}
\end{figure}

Hybrid Image Transformation (HIT) is a method proposed in~\cite{Vu}. 
To construct a HIT image, we combine the red and blue channels from the 
Entropy method discussed in Section~\ref{subsect:entropy}
with the green channel from the Byteclass method presented in Section~\ref{subsect:byteclass}. 
Since these occupy different color channels, we can seamlessly merge these two methods. 
Examples of HIT images from  our dataset appear in Figure~\ref{fig:viz_hit}.
Note that HIT images are constructed from techniques that use a Hilbert curve byte layout.

\subsubsection{Byte Bigrams}\label{subsect:bigrams}

\begin{figure}[!htb]
    \centering
        \includegraphics[width=0.475\textwidth]{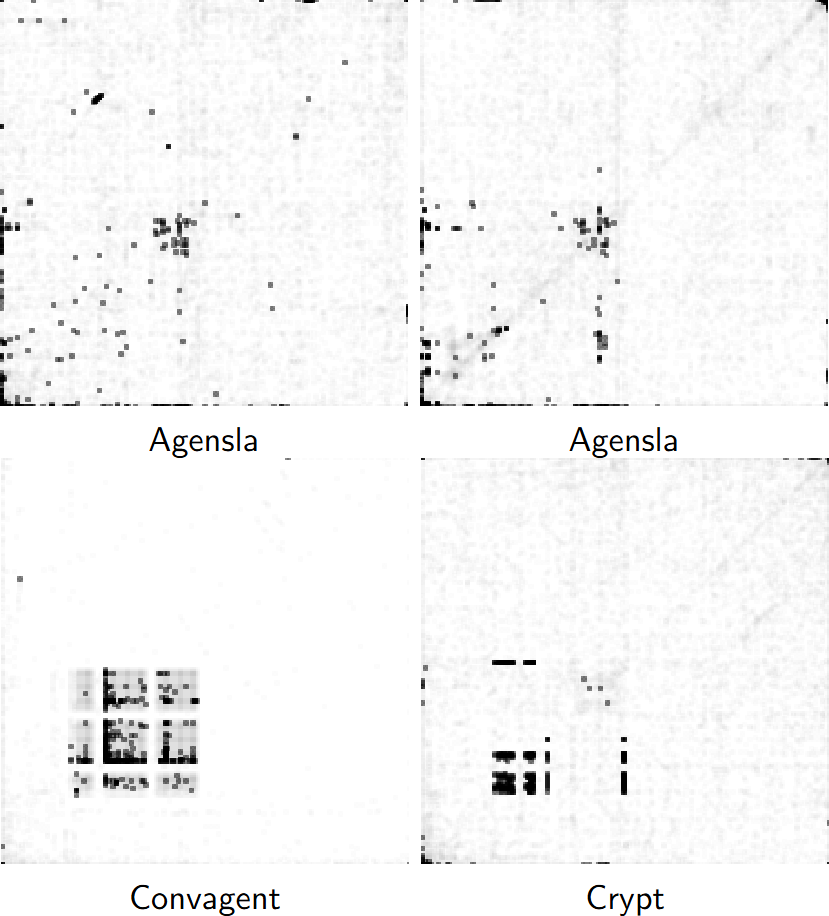}
        \caption{Cartesian bigram images of Agensla, Convagent, and Crypt}
        \label{fig:viz_bigram_cart}
\end{figure}

\begin{figure}[!htb]        
    \centering
        \includegraphics[width=0.475\textwidth]{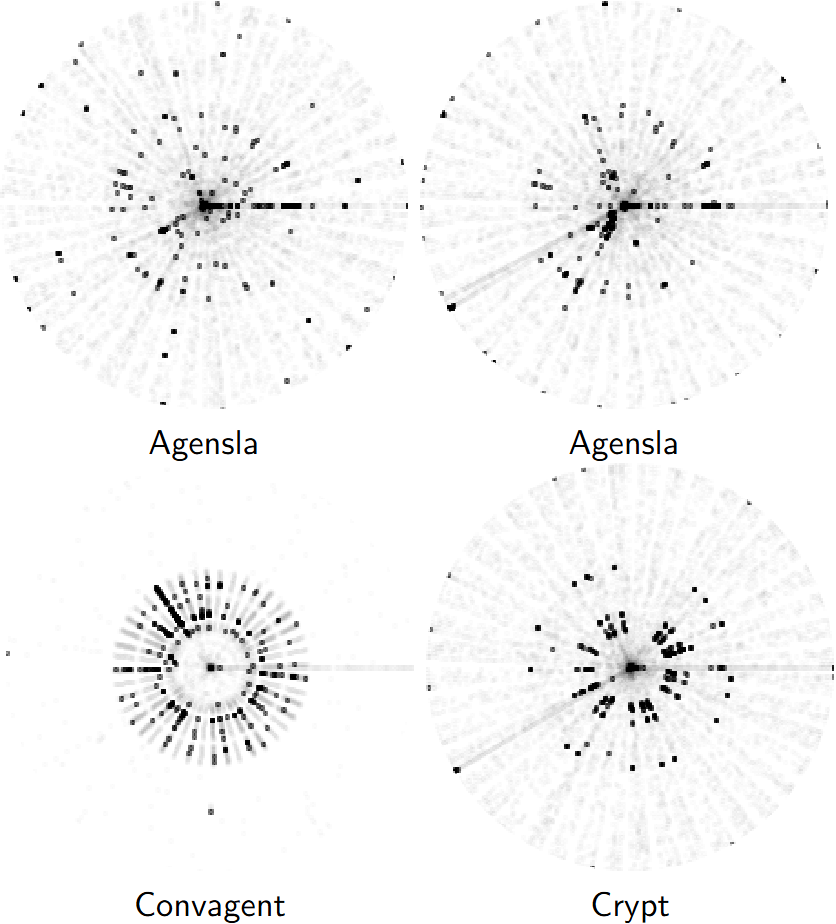}
        \caption{Polar bigram images of Agensla, Convagent, and Crypt}
        \label{fig:viz_bigram_polar}
\end{figure}

We consider two closely related bigram-based visualization methods, which we refer to as
Cartesian and Polar~\cite{Varela_2017}. Consider 
a sliding window of two consecutive bytes (or bigrams), with the pair denoted as~$(x,y)$. 
In the Cartesian approach, the pair~$(x,y)$ represents a point in the plane, while for the 
Polar approach, $x$ represents the radius and~$y$ represents the angle. Each time 
a bigram is repeated, the intensity
of the corresponding pixel in the image is increased
Examples of Cartesian bigram images from our dataset appear in Figure~\ref{fig:viz_bigram_cart},
while examples of Polar bigram images are given in Figure~\ref{fig:viz_bigram_polar}.
%As with all of our images conversion techniques, these images are ultimately
%interpreted to be of size~$224\times 224$.

\subsubsection{Spiral}\label{subsect:spiral}

%%%%% Strange to use histograms here, as the intention was to consider
%%%%% raw byte values in a spiral--like a spiral version of Grayscale images
For this technique, we were inspired by the paper~\cite{RUSTNGUYEN2023103098}, 
where a spiral visualization yields strong results in a different problem domain. 
The Spiral approach
discussed here can be viewed as a simplified form of the Hilbert curve technique,
but based on histograms, instead of raw byte values.

Each sample in the RawMalTF dataset includes a feature vector of length~256, 
consisting of a normalized histogram of byte values~\cite{2025rawmaltfrawmalwaredataset}. 
%%%%% Not clear how the RF is trained and how Gini value are computed?????
For all dataset samples, we consider each byte
position as a feature, and use Random Forest to compute the Gini value of each feature. 
Ranking from highest to lowest Gini importance, we obtain an order of importance for each byte. 
We then normalize the value of each feature across all samples, which results in values 
between~0 and~1.

\begin{figure}[!htb]
	\centering
        \includegraphics[width=0.5\textwidth]{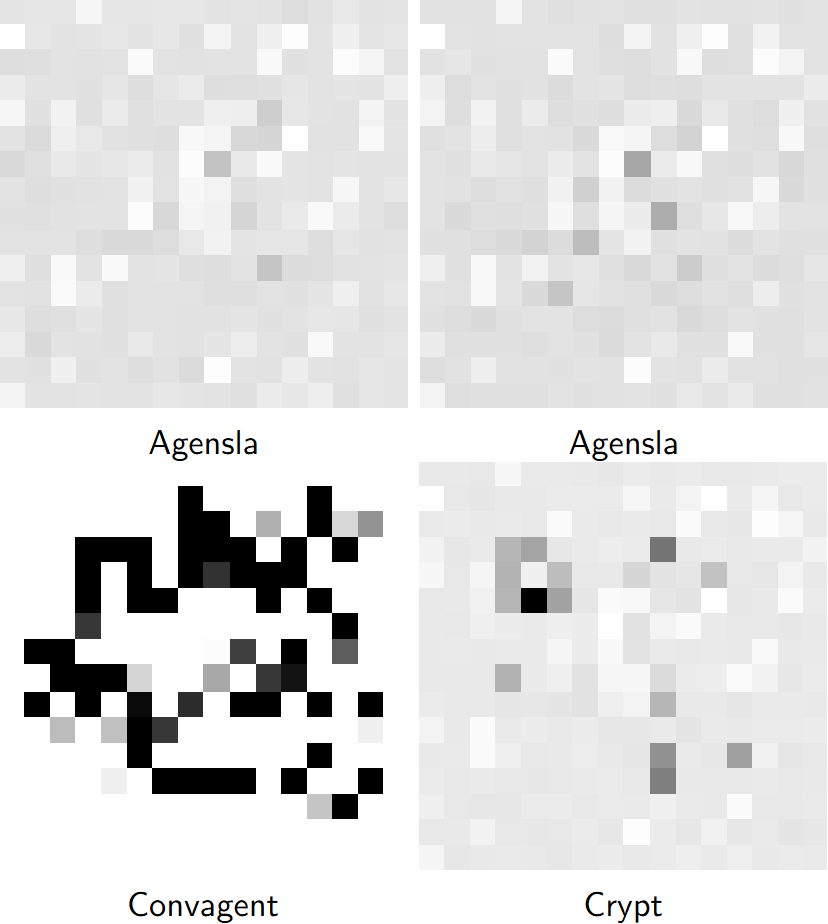}
        \caption{Spiral images of Agensla, Convagent, and Crypt}
        \label{fig:viz_spiral}
\end{figure}

To visualize a sample as a spiral, we use the normalized byte histogram values, 
re-ordered based on the Gini values as discussed above. We then multiply the normalized 
values by~255 and truncate to obtain values ranging from~0 to~255. This range of values 
represent the luminosity of a box, with~0 being white and~255 being black. 
We spiral from the center to generate a~$16\times 16$ array from these~256 byte values. 
As a final step, we use \texttt{matplotlib}~\cite{matplotlib} 
to export the results as a~$224\times 224$ image. 
Examples of spiral images from our dataset appear in Figure~\ref{fig:viz_spiral}.

Note that whereas all of the other image generation
approaches considered in this paper
are based on raw bytes, the Spiral technique discussed here
is based on histograms.
Consequently, these Spiral images are  
more directly comparable to the baseline histogram features than the other image
generation techniques.

\section{Experimental Setup}\label{sect:experimentprep}

In this section, we provide some details on our experimental design and evaluation. 
Specifically, we focus on data preparation, hyperparameter tuning, and 
evaluation metrics.

\subsection{Dataset Preparation}

As mentioned above, we randomly select~1000 samples from each of~17 
malware families obtained from the RawMalTF dataset.
For each of these~17,000 samples, we have a normalized histogram,
and we generate the eight images
described in Section~\ref{subsect:imagetransformations}. The
resulting data is split into train, validation, and test sets for multiclass classification. 
For all experiments, the train:test:validation ratios are~80:10:10.

\subsection{Hyperparameter Tuning}

For most of our models, we use 
Optuna~\cite{akiba2019optunanextgenerationhyperparameteroptimization}
for hyperparameter tuning. 
Optuna is an open-source framework that employs efficient sampling algorithms 
to optimize the values of hyperparameters. For each trial, Optuna starts with proposed 
parameter values, evaluates the objective function,
and updates its internal model of the sample space. 
To ensure reproducibility, we set the random seed to~42 
during model initialization. 
The CNN models (VGG16, InceptionV3, DenseNet121) 
took much longer to train, so for each of these models,
we tested a small number of hyperparameter values via a grid search. 

We have listed the hyperparameters tested in Table~\ref{tab:hype}, 
where~$S$ denotes the number of training samples, 
and all numerical value ranges written as ``$X$ to $Y$'' indicate that 
Optuna sampled values in the range from $X$ to~$Y$, inclusive.

\begin{table}[!htb]
\caption{Hyperparameters tested}\label{tab:hype}
\centering
\adjustbox{scale=0.785}{
\begin{tabular}{ccc}
\toprule
\textbf{Classifier} & \textbf{Hyperparameter} & \textbf{Values tested} \\ \midrule
    \multirow{5}{*}{KNN} & \texttt{k} & \{1 to $\lfloor\sqrt{S}\rfloor$\} \\
        & \texttt{weights} & \{\texttt{uniform}, \texttt{distance}\} \\
        & \texttt{metric} &  \{\texttt{euclidean}, \texttt{manhattan}, \texttt{minkowski}\} \\
        & \texttt{p} (minkowski) & \{1, 3\} \\
        & \texttt{p} (non-minkowski) & 2 \\ \midrule
    \multirow{7}{*}{MLP} & \texttt{hidden\_layer\_size} & (100,)\\
        & \texttt{activation} & \{\texttt{relu}, \texttt{tanh}, \texttt{logistic}\}\\
        & \texttt{learning\_rate\_init} & \{0.0001 to 0.1\}\\ 
        & \texttt{batch\_size} & \{32, 64, 128\}\\
        & \texttt{max\_iter} & \{100, 300\}\\
        & \texttt{alpha} & \{0.00001 to 0.1\}\\
        & \texttt{solver} & \{\texttt{adam}, \texttt{sgd}\}\\ \midrule
    \multirow{6}{*}{SVM} & \texttt{kernel} & \{\texttt{rbf}\}\\
        & \texttt{C} & \{0.1 to 100\}\\
        & \texttt{gamma} & \{0.0001 to 1.0\}\\
        & \texttt{cache\_size} & 1000\\
        & \texttt{max\_iter} & 1000\\
        & \texttt{tol} & 0.001\\ \midrule
    \multirow{5}{*}{XGBoost} & \texttt{max\_depth} & \{3 to 10\}\\
        & \texttt{learning\_rate} & \{0.01 to 0.3\}\\
        & \texttt{n\_estimators} & \{50 to 500\}\\
        & \texttt{subsample} & \{0.6 to 1.0\}\\
        & \texttt{colsample\_bytree} & \{0.6 to 1.0\}\\ \midrule
    \multirow{3}{*}{VGG16} & \texttt{learning\_rate} & \{0.0001, 0.001, 0.01\}\\
        & \texttt{momentum} & \{0.9, 0.99\}\\
        & \texttt{optimizer} & \texttt{sgd}\\ \midrule
    \multirow{3}{*}{InceptionV3} & \texttt{learning\_rate} & \{0.0001, 0.001, 0.01\}\\
        & \texttt{momentum} & \{0.9, 0.99\}\\
        & \texttt{optimizer} & \texttt{sgd}\\ \midrule
    \multirow{3}{*}{DenseNet121} & \texttt{learning\_rate} & \{0.0001, 0.001, 0.01\}\\
        & \texttt{momentum} & \{0.9, 0.99\}\\
        & \texttt{optimizer} & \texttt{sgd}\\
        \bottomrule
\end{tabular}
}
\end{table}

\subsection{Evaluation Metrics}

To evaluate our results on the test set, 
we use accuracy, precision, recall, and F1-score as metrics.
Of course, accuracy is simply the fraction of samples that are
correctly classified. Let~$\tp$ be the number of true positives,
$\fp$ be the number of false positives, and $\fn$ be the number of false negatives.
Then 
$$
  \precision = \frac{\tp}{\tp + \fp},\ \ 
  \recall = \frac{\tp}{\tp + \fn},
    \mbox{\ \ and\ \ }
  \mbox{F1-score} = 2\cdot\frac{\precision\cdot\recall}{\precision + \recall}
$$
%Note that the F1-score is computed as the harmonic mean of the precision and recall.  
Since our datasets are balanced, the accuracy and recall
will be identical in all cases.

\section{Experimental Results}\label{sect:experimentresults}

Tables~\ref{tab:base} through~\ref{tab:biPolar} in Appendix~A
contain detailed results---in terms of accuracy, precision, recall, and F1-score---for 
all of our experiments. Figures~\ref{fig:conf_KB} through~\ref{fig:conf_KHP} 
in Appendix~B provide confusion matrices for selected experiments. 
Based on the confusion matrices in Appendix~B, we note that 
\DCRat\ is the easiest family to classify, with \Makoob\ generally being
the next easiest, while the most difficult families to classify varies,
depending on the features and classification technique employed.

In the remainder of this section, we 
provide graphs that highlight various aspects of the experiments.
We also discuss the significance of these results.

%\begin{figure}[!htb]
%\centering
%\input figures/bar_grayscale.tex
%\caption{Grayscale results}\label{fig:gray}
%\end{figure}

\subsection{Baseline Results}

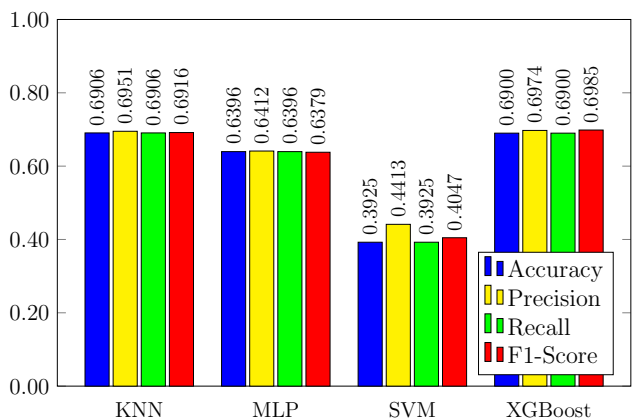
\begin{figure}[!htb]
\centering
\adjustbox{scale=0.95}{
\begin{tikzpicture}[scale=0.8, every node/.style={scale=1.0}]
\pgfkeys{/pgf/number format/.cd,1000 sep={}}
\begin{axis}[%axis x line*=bottom,
	%bar shift=0pt,
        width  = 0.75*\textwidth,
        height = 8.0cm,
        ymin=0.0,ymax=1.0,
        ytick={0.0, 0.2, 0.4, 0.6, 0.8, 1.0},
        major x tick style = transparent,
        ybar=5*\pgflinewidth,
        bar width=12.0pt,
%        ymajorgrids = true,
%        xlabel = {Learning technique},
%        ylabel = {Accuracy},
        ylabel style = {scale = 0.95},
%        symbolic x coords={MLR, SVC, MLP, CNN, RNN, LSTM, RF, XGBoost},
%        xticklabels={MLR, SVC, MLP, CNN, RNN, LSTM, RF, XGBoost},
        symbolic x coords={A, B, C, D},
        xticklabels={KNN, MLP, SVM, XGBoost},
	y tick label style={scale=0.95,
    		/pgf/number format/.cd,
   		fixed,
   		fixed zerofill,
%		sep=,
    		precision=2},
%	yticklabel pos=right,
        xtick = data,
        x tick label style={scale=0.9,
 %       		rotate=60,
%		font=\small,
%		anchor=north east,
%		inner sep=0mm
		},
%		font=\small},
%        scaled y ticks = false,
	%%%%% numbers on bars and rotated
        nodes near coords,
        every node near coord/.append style={rotate=90, scale=0.875,
        								   anchor=west, 
								   /pgf/number format/.cd,
								   fixed,
								   fixed zerofill,
%								   sep=,
								   precision=4},
        %%%%%
%        enlarge x limits=0.03,
%        enlarge x limits=0.06,
        enlarge x limits=0.20,
        legend cell align=left,
        legend pos=south east,
%        legend style={
%%                at={(1,1.05)},
%%                anchor=south east,
%%	        nodes={rotate=90},%%%%% rotate text in legend
%%                at={(0.125,0)},
%%                at={(0.125,0)},
%%                at={(0.8775,0)},
%                at={(0.82,0.56)},
%                anchor=south,
%                column sep=1ex
%        },
]
\addplot [fill=blue,opacity=1.00]
coordinates {
(A, 0.6906)
(B, 0.6396)
(C, 0.3925)
(D, 0.6900)
};
\addlegendentry{Accuracy}
\addplot [fill=yellow,opacity=1.00]
coordinates {
(A, 0.6951)
(B, 0.6412)
(C, 0.4413)
(D, 0.6974)
};
\addlegendentry{Precision}
\addplot [fill=green,opacity=1.00]
coordinates {
(A, 0.6906)
(B, 0.6396)
(C, 0.3925)
(D, 0.6900)
};
\addlegendentry{Recall}
\addplot [fill=red,opacity=1.00]
coordinates {
(A, 0.6916)
(B, 0.6379)
(C, 0.4047)
(D, 0.6985)
};
\addlegendentry{F1-Score}
\end{axis}
\end{tikzpicture}
}
\caption{Baseline results (byte histogram features)}\label{fig:base}
\end{figure}

For our baseline experiments, we tested all non-CNN models
using byte histogram features. The results of these experiments
appear in the form of a bar graph in
Figure~\ref{fig:base}. Note that the best accuracy that we attain 
is~0.6906 using KNN, while XGBoost performs almost as well,
and SVM gives significantly worse results. 
These results indicate that distinguishing between
the~17 classes in our dataset is indeed a challenging problem.

\subsection{Image-Based Classification Experiments}

Next, we compare the accuracies achieved for each of the eight 
distinct image conversion techniques under consideration, namely,
Grayscale, HIT, Entropy, Byteclass, Hilbert, Spiral, Cartesian, 
and Polar---see Section~\ref{subsect:imagetransformations} for details on these
malware-to-image conversion techniques.

For each of the eight image conversion techniques, 
we train and test each of the seven models discussed in Section~\ref{sect:models},
namely, KNN, MLP, SVM, XGBoost, VGG16, InceptionV3, and DenseNet121.
Recall that while VGG16, InceptionV3, and DenseNet121 are trained directly on the images,
for KNN, MLP, SVM, and XGBoost, we consider two distinct cases---one using 
features based on HOG and one using Haralick features. Overall, 
this gives us a total of~88 image-based experiments, in addition
to the~4 baseline experiments, for grand total of~92 distinct experiments.

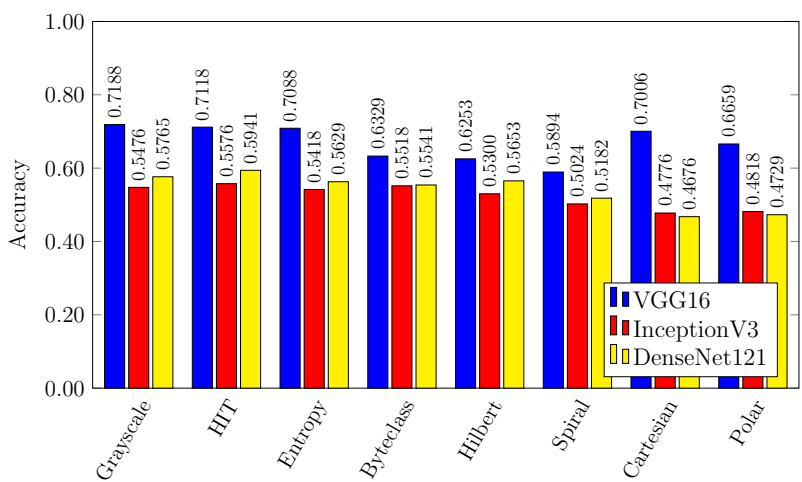
\begin{figure}[!htb]
\centering
\adjustbox{scale=0.95}{
\begin{tikzpicture}[scale=0.8, every node/.style={scale=1.0}]
\pgfkeys{/pgf/number format/.cd,1000 sep={}}
\begin{axis}[%axis x line*=bottom,
	%bar shift=0pt,
        width  = 0.9*\textwidth,
        height = 8.0cm,
        ymin=0.0,ymax=1.0,
        ytick={0.0, 0.2, 0.4, 0.6, 0.8, 1.0},
        major x tick style = transparent,
        ybar=5*\pgflinewidth,
        bar width=10.0pt,
%        ymajorgrids = true,
%        xlabel = {Learning technique},
        ylabel = {Accuracy},
        ylabel style = {scale = 0.95},
%        symbolic x coords={MLR, SVC, MLP, CNN, RNN, LSTM, RF, XGBoost},
%        xticklabels={MLR, SVC, MLP, CNN, RNN, LSTM, RF, XGBoost},
        symbolic x coords={A, B, C, D, E, F, G, H},
        xticklabels={Grayscale, HIT, Entropy, Byteclass, Hilbert, Spiral, Cartesian, Polar},
	y tick label style={scale=0.95,
    		/pgf/number format/.cd,
   		fixed,
   		fixed zerofill,
%		sep=,
    		precision=2},
%	yticklabel pos=right,
        xtick = data,
        x tick label style={scale=0.90,
        		rotate=60,
		anchor=north east,
		inner sep=0mm
		},
%        scaled y ticks = false,
	%%%%% numbers on bars and rotated
        nodes near coords,
        every node near coord/.append style={rotate=90, scale=0.8,
        								   anchor=west, 
								   /pgf/number format/.cd,
								   fixed,
								   fixed zerofill,
%								   sep=,
								   precision=4},
        %%%%%
%        enlarge x limits=0.03,
%        enlarge x limits=0.06,
        enlarge x limits=0.075,
        legend cell align=left,
        legend pos=south east,
%        legend style={
%%                at={(1,1.05)},
%%                anchor=south east,
%%	        nodes={rotate=90},%%%%% rotate text in legend
%%                at={(0.125,0)},
%%                at={(0.125,0)},
%%                at={(0.8775,0)},
%                at={(0.82,0.56)},
%                anchor=south,
%                column sep=1ex
%        },
]
\addplot [fill=blue,opacity=1.00]
coordinates {
(A, 0.7188)
(B, 0.7118)
(C, 0.7088)
(D, 0.6329)
(E, 0.6253)
(F, 0.5894)
(G, 0.7006)
(H, 0.6659)
};
\addlegendentry{VGG16}
\addplot [fill=red,opacity=1.00]
coordinates {
(A, 0.5476)
(B, 0.5576)
(C, 0.5418)
(D, 0.5518)
(E, 0.5300)
(F, 0.5024)
(G, 0.4776)
(H, 0.4818)
};
\addlegendentry{InceptionV3}
\addplot [fill=yellow,opacity=1.00]
coordinates {
(A, 0.5765)
(B, 0.5941)
(C, 0.5629)
(D, 0.5541)
(E, 0.5653)
(F, 0.5182)
(G, 0.4676)
(H, 0.4729)
};
\addlegendentry{DenseNet121}
\end{axis}
\end{tikzpicture}
}
\caption{Accuracy of pre-trained image-based models}\label{fig:acc_pre}
\end{figure}

From the bar graph in Figure~\ref{fig:acc_pre}, we observe that VGG16 substantially
outperforms the other two pre-trained image-based models---InceptionV3 and DenseNet121---across 
all image conversion techniques tested. VGG16 is noted for being relatively lightweight, in
the sense of requiring less training data than most other pre-trained models. Hence, it is
conceivable that InceptionV3 and DenseNet121 results could be improved using a larger
dataset. With respect to VGG16, we see that Grayscale and HIT
yield the best results, while Entropy and Cartesian give results that are comparable,
while Spiral performs the worst.

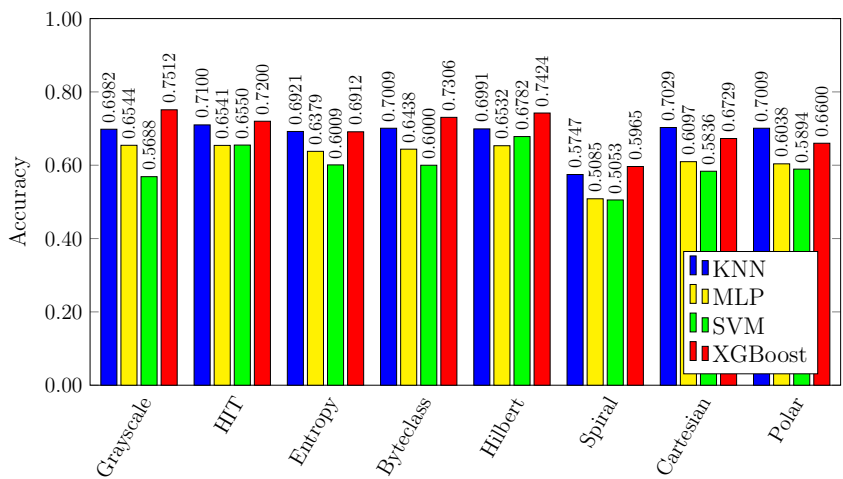
\begin{figure}[!htb]
\centering
\adjustbox{scale=0.95}{
\begin{tikzpicture}[scale=0.8, every node/.style={scale=1.0}]
\pgfkeys{/pgf/number format/.cd,1000 sep={}}
\begin{axis}[%axis x line*=bottom,
	%bar shift=0pt,
        width  = 0.95*\textwidth,
        height = 8.0cm,
        ymin=0.0,ymax=1.0,
        ytick={0.0, 0.2, 0.4, 0.6, 0.8, 1.0},
        major x tick style = transparent,
        ybar=5*\pgflinewidth,
        bar width=8.0pt,
%        ymajorgrids = true,
%        xlabel = {Learning technique},
        ylabel = {Accuracy},
        ylabel style = {scale = 0.95},
%        symbolic x coords={MLR, SVC, MLP, CNN, RNN, LSTM, RF, XGBoost},
%        xticklabels={MLR, SVC, MLP, CNN, RNN, LSTM, RF, XGBoost},
        symbolic x coords={A, B, C, D, E, F, G, H},
        xticklabels={Grayscale, HIT, Entropy, Byteclass, Hilbert, Spiral, Cartesian, Polar},
	y tick label style={scale=0.90,
    		/pgf/number format/.cd,
   		fixed,
   		fixed zerofill,
%		sep=,
    		precision=2},
%	yticklabel pos=right,
        xtick = data,
        x tick label style={scale=0.90,
        		rotate=60,
		anchor=north east,
		inner sep=0mm
		},
%		font=\small},
%        scaled y ticks = false,
	%%%%% numbers on bars and rotated
        nodes near coords,
        every node near coord/.append style={rotate=90, scale=0.775,
        								   anchor=west, 
								   /pgf/number format/.cd,
								   fixed,
								   fixed zerofill,
%								   sep=,
								   precision=4},
        %%%%%
%        enlarge x limits=0.03,
%        enlarge x limits=0.06,
        enlarge x limits=0.075,
        legend cell align=left,
        legend pos=south east,
%        legend style={
%%                at={(1,1.05)},
%%                anchor=south east,
%%	        nodes={rotate=90},%%%%% rotate text in legend
%%                at={(0.125,0)},
%%                at={(0.125,0)},
%%                at={(0.8775,0)},
%                at={(0.82,0.56)},
%                anchor=south,
%                column sep=1ex
%        },
]
\addplot [fill=blue,opacity=1.00]
coordinates {
(A, 0.6982)
(B, 0.7100)
(C, 0.6921)
(D, 0.7009)
(E, 0.6991)
(F, 0.5747)
(G, 0.7029)
(H, 0.7009)
};
\addlegendentry{KNN}
\addplot [fill=yellow,opacity=1.00]
coordinates {
(A, 0.6544)
(B, 0.6541)
(C, 0.6379)
(D, 0.6438)
(E, 0.6532)
(F, 0.5085)
(G, 0.6097)
(H, 0.6038)
};
\addlegendentry{MLP}
\addplot [fill=green,opacity=1.00]
coordinates {
(A, 0.5688)
(B, 0.6550)
(C, 0.6009)
(D, 0.6000)
(E, 0.6782)
(F, 0.5053)
(G, 0.5836)
(H, 0.5894)
};
\addlegendentry{SVM}
\addplot [fill=red,opacity=1.00]
coordinates {
(A, 0.7512)
(B, 0.7200)
(C, 0.6912)
(D, 0.7306)
(E, 0.7424)
(F, 0.5965)
(G, 0.6729)
(H, 0.6600)
};
\addlegendentry{XGBoost}
\end{axis}
\end{tikzpicture}
}
\caption{Accuracy using HOG features}\label{fig:acc_hog}
\end{figure}

Figure~\ref{fig:acc_hog} shows that for the HOG features, XGBoost and KNN
perform best, with XGBoost being better for six of the eight image conversion
techniques, while KNN is better for the remaining two image conversion 
techniques. With respect to XGBoost, Grayscale
is the best image conversion approach, 
followed in order by Hilbert, Byteclass, and HIT

\begin{figure}[!htb]
\centering
\adjustbox{scale=0.95}{
\begin{tikzpicture}[scale=0.8, every node/.style={scale=1.0}]
\pgfkeys{/pgf/number format/.cd,1000 sep={}}
\begin{axis}[%axis x line*=bottom,
	%bar shift=0pt,
        width  = 0.95*\textwidth,
        height = 8.0cm,
        ymin=0.0,ymax=1.0,
        ytick={0.0, 0.2, 0.4, 0.6, 0.8, 1.0},
        major x tick style = transparent,
        ybar=5*\pgflinewidth,
        bar width=8.0pt,
%        ymajorgrids = true,
%        xlabel = {Learning technique},
        ylabel = {Accuracy},
        ylabel style = {scale = 0.95},
%        symbolic x coords={MLR, SVC, MLP, CNN, RNN, LSTM, RF, XGBoost},
%        xticklabels={MLR, SVC, MLP, CNN, RNN, LSTM, RF, XGBoost},
        symbolic x coords={A, B, C, D, E, F, G, H},
        xticklabels={Grayscale, HIT, Entropy, Byteclass, Hilbert, Spiral, Cartesian, Polar},
	y tick label style={scale=0.90,
    		/pgf/number format/.cd,
   		fixed,
   		fixed zerofill,
%		sep=,
    		precision=2},
%	yticklabel pos=right,
        xtick = data,
        x tick label style={scale=0.90,
        		rotate=60,
		anchor=north east,
		inner sep=0mm
		},
%		font=\small},
%        scaled y ticks = false,
	%%%%% numbers on bars and rotated
        nodes near coords,
        every node near coord/.append style={rotate=90, scale=0.775,
        								   anchor=west, 
								   /pgf/number format/.cd,
								   fixed,
								   fixed zerofill,
%								   sep=,
								   precision=4},
        %%%%%
%        enlarge x limits=0.03,
%        enlarge x limits=0.06,
        enlarge x limits=0.075,
        legend cell align=left,
        legend pos=south east,
%        legend style={
%%                at={(1,1.05)},
%%                anchor=south east,
%%	        nodes={rotate=90},%%%%% rotate text in legend
%%                at={(0.125,0)},
%%                at={(0.125,0)},
%%                at={(0.8775,0)},
%                at={(0.82,0.56)},
%                anchor=south,
%                column sep=1ex
%        },
]
\addplot [fill=blue,opacity=1.00]
coordinates {
(A, 0.6953)
(B, 0.6965)
(C, 0.6241)
(D, 0.7062)
(E, 0.6788)
(F, 0.4265)
(G, 0.5232)
(H, 0.5168)
};
\addlegendentry{KNN}
\addplot [fill=yellow,opacity=1.00]
coordinates {
(A, 0.4424)
(B, 0.3941)
(C, 0.3962)
(D, 0.4559)
(E, 0.4494)
(F, 0.3394)
(G, 0.4074)
(H, 0.3832)
};
\addlegendentry{MLP}
\addplot [fill=green,opacity=1.00]
coordinates {
(A, 0.6379)
(B, 0.6517)
(C, 0.5700)
(D, 0.6694)
(E, 0.6556)
(F, 0.4371)
(G, 0.5612)
(H, 0.5262)
};
\addlegendentry{SVM}
\addplot [fill=red,opacity=1.00]
coordinates {
(A, 0.7229)
(B, 0.7076)
(C, 0.6341)
(D, 0.7147)
(E, 0.7100)
(F, 0.4653)
(G, 0.5312)
(H, 0.5065)
};
\addlegendentry{XGBoost}
\end{axis}
\end{tikzpicture}
}
\caption{Accuracy using Haralick features}\label{fig:acc_har}
\end{figure}
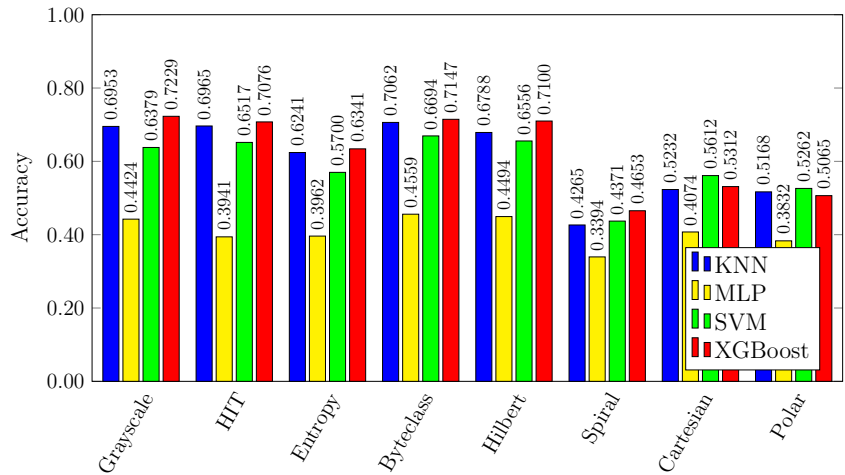

From Figure~\ref{fig:acc_har} we observe that the results using the Haralick
features are generally worse than those obtained using the HOG features, with
the MLP performing much worse than in the HOG case.
The only notable exception is that in a few cases, 
the SVM model performs better with the Haralick features. 
With respect to the image conversion techniques, Grayscale is best,
followed in order by Byteclass, Hilbert, and HIT. For the Haralick features,
Spiral, Cartesian, and Polar all yield consistently poor results.

\begin{figure}[!htb]
\centering
\adjustbox{scale=0.95}{
\begin{tikzpicture}[scale=0.8, every node/.style={scale=1.0}]
\pgfkeys{/pgf/number format/.cd,1000 sep={}}
\begin{axis}[%axis x line*=bottom,
	%bar shift=0pt,
        width  = 0.65*\textwidth,
        height = 8.0cm,
        ymin=0.0,ymax=1.0,
        ytick={0.0, 0.2, 0.4, 0.6, 0.8, 1.0},
        major x tick style = transparent,
        ybar=5*\pgflinewidth,
        bar width=16.0pt,
%        ymajorgrids = true,
%        xlabel = {Learning technique},
        ylabel = {Accuracy},
        ylabel style = {scale = 0.95},
%        symbolic x coords={MLR, SVC, MLP, CNN, RNN, LSTM, RF, XGBoost},
%        xticklabels={MLR, SVC, MLP, CNN, RNN, LSTM, RF, XGBoost},
        symbolic x coords={A, B, C, D, E, F, G, H, I},
        xticklabels={Baseline, 
                           Grayscale, 
                           HIT, 
                           Entropy, 
                           Byteclass, 
                           Hilbert, 
                           Spiral, 
                           Cartesian, 
                           Polar},
	y tick label style={scale=0.90,
    		/pgf/number format/.cd,
   		fixed,
   		fixed zerofill,
%		sep=,
    		precision=2},
%	yticklabel pos=right,
        xtick = data,
        x tick label style={scale=0.90,
        		rotate=60,
		anchor=north east,
		inner sep=0mm
		},
%        scaled y ticks = false,
	%%%%% numbers on bars and rotated
%        nodes near coords,
        nodes near coords={\pgfmathfloatifflags{\pgfplotspointmeta}{0}{}{\pgfmathprintnumber{\pgfplotspointmeta}}},
        every node near coord/.append style={rotate=90, scale=0.9250,
        								   anchor=west, 
								   /pgf/number format/.cd,
								   fixed,
								   fixed zerofill,
%								   sep=,
								   precision=4},
        %%%%%
%        enlarge x limits=0.03,
%        enlarge x limits=0.06,
        enlarge x limits=0.10,
        legend cell align=left,
        legend pos=south east,
        every axis plot/.append style={
          ybar,
          bar shift=0pt,
          fill
        }
%        legend style={
%%                at={(1,1.05)},
%%                anchor=south east,
%%	        nodes={rotate=90},%%%%% rotate text in legend
%%                at={(0.125,0)},
%%                at={(0.125,0)},
%%                at={(0.8775,0)},
%                at={(0.82,0.56)},
%                anchor=south,
%                column sep=1ex
%        },
]
\addplot [black,thick,fill=red,opacity=1.00]
coordinates {
(A, 0.6906)
(B, 0.0)
(C, 0.0)
(D, 0.0)
(E, 0.0)
(F, 0.0)
(G, 0.0)
(H, 0.0)
(I, 0.0)
};
\addplot [black,thick,fill=blue,opacity=1.00]
coordinates {
(B, 0.7512)
(C, 0.7200)
(D, 0.7088)
(E, 0.7306)
(F, 0.7424)
(G, 0.5965)
(H, 0.7029)
(I, 0.7009)
};
%\addplot [fill=blue,opacity=1.00]
%coordinates {
%(A, 0.5172)
%(B, 0.7512)
%(C, 0.7200)
%(D, 0.7088)
%(E, 0.7306)
%(F, 0.7424)
%(G, 0.5965)
%(H, 0.7029)
%(I, 0.7009)
%};
\end{axis}
\end{tikzpicture}
}
\caption{Highest accuracy for baseline and each image conversion technique}\label{fig:acc_best}
\end{figure}
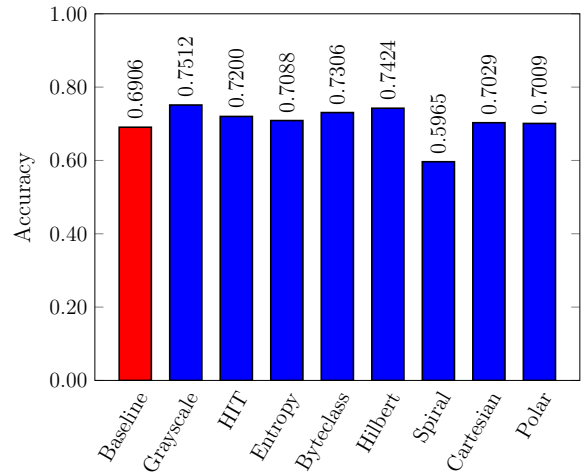

In Figure~\ref{fig:acc_best} we compare the best accuracy for the baseline
cases to the best accuracy for each of the image conversion techniques.
From these results, we observe that the baseline models trained on 
histogram features underperform all image-based
cases, with the exception of the Spiral image type. 
Among the image conversion techniques, Grayscale yields the
best results, followed closely in order by Hilbert, Byteclass, and HIT, 
with Entropy, Cartesian, and Polar performing slightly worse. 
%Only the Spiral technique yields result that fail to be within about~5\%\ of the best case.

\begin{figure}[!htb]
\centering
\adjustbox{scale=0.95}{
%\begin{tikzpicture}[scale=0.6, every node/.style={scale=0.8}]
\begin{tikzpicture}
	\pgfplotstableread[col sep=comma]{data/data2.csv}\csvdata
	% Boxplot groups columns, but we want rows
	\pgfplotstabletranspose\datatransposed{\csvdata} 
	\begin{axis}[
		boxplot/draw direction = y,
		boxplot/every whisker/.style={blue,thick},
		boxplot/every median/.style={blue,thick},
		boxplot/every box/.style={blue,very thick,fill=white,fill opacity=0},
		x axis line style = {opacity=0},
%		x axis line style = {opacity=1.0},
		axis x line* = bottom,
		axis y line = left,
		enlarge y limits,
		ymajorgrids,
		ymin=0.275, ymax=0.8,
		xtick = {1, 2, 3, 4, 5, 6, 7, 8},
		ylabel style = {scale = 0.80},
	        x tick label style={scale=0.75,
        			rotate=60,
			anchor=north east,
			inner sep=0mm
		},
		y tick label style={scale=0.75,
	    		/pgf/number format/.cd,
   			fixed,
   			fixed zerofill,
%			sep=,
    			precision=2},
		xticklabels = {Grayscale, HIT, Entropy, Byteclass, Hilbert, Spiral, Cartesian, Polar},
		xtick style = {draw=none}, % Hide tick line
		ylabel = {Accuracy},
		ytick = {0.0, 0.1, 0.2, 0.3, 0.4, 0.5, 0.6, 0.7, 0.8, 0.9, 1.0}
	]
		\foreach \n in {1,...,8} {
			\addplot+[boxplot, fill, draw=black] table[y index=\n] {\datatransposed};
		}
	\end{axis}
\end{tikzpicture}
}
\caption{Box-and-whisker plots}\label{fig:whisk}
\end{figure}
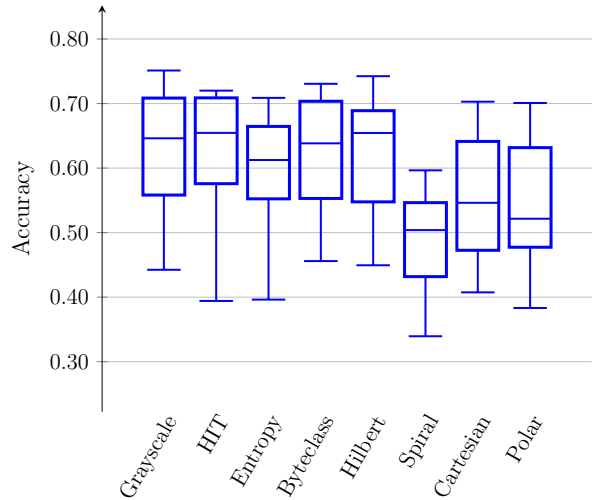

Finally, in Figure~\ref{fig:whisk} we give box-and-whisker plots 
of the accuracy of models trained using
the eight distinct image conversion techniques considered.
Each box-and-whisker plot is based on the eleven learning models tested,
consisting of the three pre-trained models, namely, VGG16, InceptionV3, and DenseNet121,
as well as the four models trained on HOG features and the four trained on Haralick
features---KNN, MLP, SVM, and XGBoost for both of these cases. 

Based on Figure~\ref{fig:whisk}, we observe that among the four best image conversion 
techniques (Grayscale, HIT, Byteclass, and Hilbert), HIT is in some sense the least
stable, due to more poor-performing models. In contrast, the box-and-whisker plots
for Grayscale, Byteclass, and Hilbert are similar to each other.

The Spiral images generally yield the worst results of all of the
image conversion techniques. Since the Spiral images are based on 
histogram values, as opposed to raw byte values, 
it is perhaps more reasonable to evaluate them in comparison
to the baseline cases, although they still fall short in that comparison. 

\subsection{Discussion}\label{sect:disc}

Overall, the experimental results presented in this section serve to emphasize 
the value of image-based techniques for malware analysis. Furthermore, the results
indicate that any number of image conversion strategies will likely yield measurable
improvement, as compared to models trained on non-image features. This provides
evidence that the image conversion process itself is of significant value in malware
analysis, with the specific image conversion technique considered playing
a secondary role. 
This helps to explain why research involving image-based 
techniques consistently shows strong results, in spite of the image conversion 
techniques often being ad hoc and poorly-motivated.

\section{Conclusion}\label{sect:conclusion}

Image-based techniques have yielded impressive results in the field of 
malware analysis. This is, perhaps, somewhat surprising, given that 
executable files naturally have a one-dimensional structure, rather than
the higher-dimensional structure of images. Furthermore, there is no
intuitively obvious ``best'' way to convert an executable into an image.
In fact, many seemingly ad hoc methods for executable-to-image
conversion have appeared in the literature, and many of these
yield excellent results.

In this paper, we compared eight distinct malware-to-image conversion 
techniques, using a variety of classifiers and features. 
Of these eight techniques---which we refer
to as Grayscale, HIT, Entropy, Byteclass, Hilbert, Spiral, Cartesian, and Polar---we found
that the simplest (i.e., Grayscale) yielded the best results. However, several of 
these techniques produced comparable results, 
which provides evidence that image analysis techniques
themselves may be the key to the success of image-based malware 
research, as opposed to any one specific image conversion technique.

In the realm of future work, larger-scale experiments could be conducted, 
involving more data, more image conversion techniques, and more classifiers.
Since the number of image conversion techniques is essentially unlimited, 
it would be useful to devise a way to categorize such techniques, and thereby
focus future research on representative examples from 
such categories. In addition, a better understanding of the reasons
for the success of image-based analysis of malware would be useful,
as it is conceivable that such an understanding could result in improved
techniques for malware analysis.

\bibliographystyle{plain}
\bibliography{refs.bib}

\section*{Appendix A}\label{app:a}

\titleformat{\section}{\normalfont\large\bfseries}{}{0em}{#1\ \thesection}
\setcounter{section}{0}
\renewcommand{\thesection}{\Alph{section}}
\renewcommand{\thesubsection}{A.\arabic{subsection}}
\setcounter{table}{0}
\renewcommand{\thetable}{A.\arabic{table}}
\setcounter{figure}{0}
\renewcommand{\thefigure}{A.\arabic{figure}}
%\subsubsection{Appendix}\label{app:a}

In this appendix, we provide detailed results for each experiment we have conducted.
Table~\ref{tab:base} contains the results of our baseline experiments.
Tables~\ref{tab:gray} through~\ref{tab:biPolar} provide the results for all of
our image-based experiments, with each table giving the results for
one of the eight image conversion approaches that we consider.
In all of these tables, the best result in each column is boxed.

\begin{table}[!htb]
\centering
\caption{Baseline results} \label{tab:base}
\adjustbox{scale=0.85}{
\begin{tabular}{l|cccc}
\toprule
\textbf{Classifier} & \textbf{Accuracy} & \textbf{Precision} & \textbf{Recall} & \textbf{F1-Score} \\
\midrule
KNN               & \fbox{0.6906} & 0.6951 & \fbox{0.6906} & \fbox{0.6916} \\
MLP 		& 0.6396 & 0.6412 & 0.6396 & 0.6379 \\
SVM                     & 0.3925& 0.4413& 0.3925& 0.4047\\
XGBoost      &  0.6900& \fbox{0.6974} & 0.6900 & 0.6895 \\
\bottomrule
\end{tabular}%
}
\end{table}

\begin{table}[!htb]
\centering
\caption{Grayscale image results} \label{tab:gray}
\adjustbox{scale=0.85}{
\begin{tabular}{l|cccc}
\toprule
\textbf{Classifier} & \textbf{Accuracy} & \textbf{Precision} & \textbf{Recall} & \textbf{F1-Score} \\
\midrule
KNN (HOG) & 0.6982& 0.7016& 0.6982& 0.6979\\
KNN (Haralick)           & 0.6953& 0.6934& 0.6953& 0.6931\\
MLP (HOG) & 0.6544 & 0.6550 & 0.6544 & 0.6537 \\
MLP (Haralick)           & 0.4424 & 0.4248 & 0.4424 & 0.4115 \\
SVM (HOG) & 0.5688& 0.5874& 0.5688& 0.5684\\
SVM (Haralick)           & 0.6379& 0.6467& 0.6379& 0.6374\\
XGBoost (HOG)           & \fbox{0.7512}& \fbox{0.7682}& \fbox{0.7512}& \fbox{0.7541}\\
XGBoost (Haralick)      & 0.7229 & 0.7270 & 0.7229 & 0.7237 \\
VGG16                   & 0.7188& 0.7256& 0.7188& 0.7183\\
InceptionV3             & 0.5476& 0.5598& 0.5476& 0.5418\\
DenseNet121             & 0.5765& 0.5782& 0.5765& 0.5745\\
\bottomrule
\end{tabular}%
}
\end{table}

\begin{table}[!htb]
\centering
\caption{HIT image results} \label{tab:hit}
\adjustbox{scale=0.85}{
\begin{tabular}{l|cccc}
\toprule
\textbf{Classifier} & \textbf{Accuracy} & \textbf{Precision} & \textbf{Recall} & \textbf{F1-Score} \\
\midrule
KNN (HOG) & 0.7100& 0.7125& 0.7100& 0.7090\\
KNN (Haralick)           & 0.6965& 0.6944& 0.6965& 0.6939\\
MLP (HOG) & 0.6541 & 0.6577 & 0.6541 & 0.6550 \\
MLP (Haralick)           & 0.3941 & 0.3757 & 0.3941 & 0.3634\\
SVM (HOG) & 0.6550& 0.6761& 0.6550& 0.6582\\
SVM (Haralick)           & 0.6517& 0.6550& 0.6517& 0.6506\\
XGBoost (HOG)           & \fbox{0.7200}&  \fbox{0.7359}& \fbox{0.7200}& \fbox{0.7216}\\
XGBoost (Haralick)      & 0.7076 & 0.7116 & 0.7076 & 0.7073 \\
VGG16                   & 0.7118& 0.7259& 0.7118& 0.7118\\
InceptionV3             & 0.5576& 0.5628& 0.5576& 0.5507\\
DenseNet121             & 0.5941& 0.5897& 0.5941& 0.5898\\
\bottomrule
\end{tabular}%
}
\end{table}

\begin{table}[!htb]
\centering
\caption{Entropy image results} \label{tab:ent}
\adjustbox{scale=0.85}{
\begin{tabular}{l|cccc}
\toprule
\textbf{Classifier} & \textbf{Accuracy} & \textbf{Precision} & \textbf{Recall} & \textbf{F1-Score} \\
\midrule
KNN (HOG)           & 0.6921& 0.7045& 0.6921& 0.6926\\
KNN (Haralick)           & 0.6241& 0.6201& 0.6241& 0.6207\\
MLP (HOG)           & 0.6379 & 0.6438 & 0.6379 & 0.6391 \\
MLP (Haralick)           & 0.3962 & 0.3664 & 0.3962 & 0.3691 \\
SVM (HOG)           & 0.6009& 0.6101& 0.6009& 0.5990\\
SVM (Haralick)           & 0.5700& 0.5737& 0.5700& 0.5685\\
XGBoost (HOG)           & 0.6912& 0.7078& 0.6912& 0.6940\\
XGBoost (Haralick)      & 0.6341& 0.6433& 0.6341&  0.6357\\
VGG16                   & \fbox{0.7088}& \fbox{0.7221}& \fbox{0.7088}& \fbox{0.7105}\\
InceptionV3             & 0.5418& 0.5795& 0.5418& 0.5447\\
DenseNet121             & 0.5629& 0.5665& 0.5629& 0.5635\\
\bottomrule
\end{tabular}%
}
\end{table}

\begin{table}[!htb]
\centering
\caption{Byteclass image results} \label{tab:bc}
\adjustbox{scale=0.85}{
\begin{tabular}{l|cccc}
\toprule
\textbf{Classifier} & \textbf{Accuracy} & \textbf{Precision} & \textbf{Recall} & \textbf{F1-Score} \\
\midrule
KNN (HOG)           & 0.7009& 0.6978& 0.7009& 0.6975\\
KNN (Haralick)           & 0.7062& 0.7050& 0.7062& 0.7027\\
MLP (HOG)           & 0.6438 & 0.6460 & 0.6438 & 0.6434 \\
MLP (Haralick)           & 0.4559 & 0.4534 & 0.4559 & 0.4427 \\
SVM (HOG)           & 0.6000& 0.6456& 0.6000& 0.6073\\
SVM (Haralick)           & 0.6694& 0.6740& 0.6694& 0.6684\\
XGBoost (HOG)           & \fbox{0.7306}& \fbox{0.7428}& \fbox{0.7306}& \fbox{0.7319}\\
XGBoost (Haralick)      & 0.7147 & 0.7167 & 0.7147 & 0.7136 \\
VGG16                   & 0.6329& 0.6638& 0.6329& 0.6326\\
InceptionV3             & 0.5518& 0.5735& 0.5518& 0.5450\\
DenseNet121             & 0.5541& 0.5516& 0.5541& 0.5446\\
\bottomrule
\end{tabular}%
}
\end{table}

\begin{table}[!htb]
\centering
\caption{Hilbert image results}\label{tab:bc2}
\adjustbox{scale=0.85}{
\begin{tabular}{l|cccc}
\toprule
\textbf{Classifier} & \textbf{Accuracy} & \textbf{Precision} & \textbf{Recall} & \textbf{F1-Score} \\
\midrule
KNN (HOG) & 0.6991& 0.6957& 0.6991& 0.6961\\
KNN (Haralick)           & 0.6788& 0.6740& 0.6788& 0.6746\\
MLP (HOG) & 0.6532 & 0.6540 & 0.6532 & 0.6532 \\
MLP (Haralick)           & 0.4494 & 0.4502 & 0.4494 & 0.4384 \\
SVM (HOG) & 0.6782& 0.6931& 0.6782& 0.6803\\
SVM (Haralick)           & 0.6556& 0.6621& 0.6556& 0.6530\\
XGBoost (HOG)           & \fbox{0.7424}& \fbox{0.7580}& \fbox{0.7424}& \fbox{0.7580}\\
XGBoost (Haralick)      & 0.7100 & 0.7151 & 0.7100 & 0.7091 \\
VGG16                   & 0.6253& 0.6415& 0.6253& 0.6218\\
InceptionV3             & 0.5300& 0.5397& 0.5300& 0.5243\\
DenseNet121             & 0.5653& 0.5623& 0.5653& 0.5623\\
\bottomrule
\end{tabular}%
}
\end{table}

\begin{table}[!htb]
\centering
\caption{Spiral image results}\label{tab:spiral}
\adjustbox{scale=0.85}{
\begin{tabular}{l|cccc}
\toprule
\textbf{Classifier} & \textbf{Accuracy} & \textbf{Precision} & \textbf{Recall} & \textbf{F1-Score} \\
\midrule
KNN (HOG) & 0.5747& 0.5677& 0.5747& 0.5666\\
KNN (Haralick)           & 0.4265& 0.4278& 0.4265& 0.4228\\
MLP (HOG) & 0.5085 & 0.5029 & 0.5085 & 0.5022 \\
MLP (Haralick)           & 0.3394 & 0.3160 & 0.3394 & 0.3147 \\
SVM (HOG) & 0.5053& 0.5312& 0.5053& 0.4978\\
SVM (Haralick)           & 0.4371& 0.4567& 0.4371& 0.4344\\
XGBoost (HOG)           & \fbox{0.5965}& 0.6055& \fbox{0.5965}& \fbox{0.5936}\\
XGBoost (Haralick)      & 0.4653 & 0.4781 & 0.4653 & 0.4660 \\
VGG16                   & 0.5894& \fbox{0.6422} & 0.5894& 0.5720\\
InceptionV3             & 0.5024& 0.4842& 0.5024& 0.4819\\
DenseNet121             & 0.5182& 0.5183& 0.5182& 0.5082\\
\bottomrule
\end{tabular}%
}
\end{table}

\begin{table}[!htb]
\centering
\caption{Cartesian bigram image results}\label{tab:biCart}
\adjustbox{scale=0.85}{
\begin{tabular}{l|cccc}
\toprule
\textbf{Classifier} & \textbf{Accuracy} & \textbf{Precision} & \textbf{Recall} & \textbf{F1-Score} \\
\midrule
KNN (HOG) & \fbox{0.7029}& 0.7026& \fbox{0.7029}&  \fbox{0.7021}\\
KNN (Haralick)           & 0.5232& 0.5197& 0.5232& 0.5189\\
MLP (HOG) & 0.6097 & 0.6092 & 0.6097 & 0.6068 \\
MLP (Haralick)           & 0.4074 & 0.4012 & 0.4074 & 0.3815 \\
SVM (HOG) & 0.5836& 0.5913& 0.5836& 0.5766\\
SVM (Haralick)           & 0.5612& 0.5658& 0.5612& 0.5601\\
XGBoost (HOG)           & 0.6729& 0.6743& 0.6729& 0.6678\\
XGBoost (Haralick)      & 0.5312 & 0.5355 & 0.5312 & 0.5275 \\
VGG16                   & 0.7006& \fbox{0.7055}& 0.7006& 0.7003\\
InceptionV3             & 0.4776& 0.5015& 0.4776& 0.4645\\
DenseNet121             & 0.4676& 0.4641& 0.4676& 0.4594\\
\bottomrule
\end{tabular}%
}
\end{table}

\begin{table}[!htb]
\centering
\caption{Polar bigram image results}\label{tab:biPolar}
\adjustbox{scale=0.85}{
\begin{tabular}{l|cccc}
\toprule
\textbf{Classifier} & \textbf{Accuracy} & \textbf{Precision} & \textbf{Recall} & \textbf{F1-Score} \\
\midrule
KNN (HOG) &  \fbox{0.7009}& \fbox{0.7020}&  \fbox{0.7009}& \fbox{0.7006}\\
KNN (Haralick)           & 0.5168& 0.5143& 0.5168& 0.5124\\
MLP (HOG) & 0.6038 & 0.6046 & 0.6038 & 0.6010 \\
MLP (Haralick)           & 0.3832 & 0.3560 & 0.3832 & 0.3516 \\
SVM (HOG) & 0.5894& 0.5962& 0.5894& 0.5832\\
SVM (Haralick)           & 0.5262& 0.5361& 0.5262& 0.5228\\
XGBoost (HOG)           & 0.6600& 0.6671& 0.6600& 0.6579\\
XGBoost (Haralick)      & 0.5065& 0.5083 & 0.5065 & 0.5033 \\
VGG16                   & 0.6659& 0.6703& 0.6659& 0.6646\\
InceptionV3             & 0.4818& 0.4986& 0.4818& 0.4737\\
DenseNet121             & 0.4729& 0.4681& 0.4729& 0.4667\\
\bottomrule
\end{tabular}%
}
\end{table}

\clearpage

\section*{Appendix B}\label{app:b}

\titleformat{\section}{\normalfont\large\bfseries}{}{0em}{#1\ \thesection}
\setcounter{section}{0}
\renewcommand{\thesection}{\Alph{section}}
\renewcommand{\thesubsection}{B.\arabic{subsection}}
\setcounter{table}{0}
\renewcommand{\thetable}{B.\arabic{table}}
\setcounter{figure}{0}
\renewcommand{\thefigure}{B.\arabic{figure}}

In this appendix, we provide confusion matrices for selected experiments.
We have conducted a large number of experiments, and it is not feasible
to provide confusion matrices for every case.

\begin{figure}[!htb]
\centering
%\begin{tikzpicture}[scale=0.8,every node/.style={scale=0.8}]
\begin{tikzpicture}[scale=0.60]
    \begin{axis}[%colorbar/width=2.5mm,
        width=16cm,
        height=16cm,
%        xlabel={\LARGE Predicted class},
%        ylabel={\LARGE Actual class},
%        colormap={blackwhite}{gray(0cm)=(1); gray(1cm)=(0.5)},
%	colormap={bluewhite}{color=(white) color=(blue)},
%	colormap={bluewhite}{color=(white) rgb255=(0,191,255)},
	colormap={bluewhite}{color=(white) rgb255=(100,149,237)},
        xticklabels={
        Agensla,
        Androm,
        Convagent,
        Crypt,
        Crysan,
        DCRat,
        Injuke,
        Makoob,
        Mokes,
        Noon,
        Remcos,
        Seraph,
        SnakeLogger,
        Stealerc,
        Strab,
        Taskun,
        Zenpak
        },
        xtick={0,...,16},
        xtick style={draw=none},
	xticklabel style={scale=1.25,anchor=east,rotate=60,yshift=-5pt,font=\tt},
        yticklabels={
        Agensla,
        Androm,
        Convagent,
        Crypt,
        Crysan,
        DCRat,
        Injuke,
        Makoob,
        Mokes,
        Noon,
        Remcos,
        Seraph,
        SnakeLogger,
        Stealerc,
        Strab,
        Taskun,
        Zenpak
        },
        ytick={0,...,16},
        ytick style={draw=none},
        enlargelimits=false,
        yticklabel style={scale=1.25,font=\tt},
        colorbar,
        colorbar style={
%     	  	width=0.05*\pgfkeysvalueof{/pgfplots/parent axis width},%%% added this
%     	  	height=0.5*\pgfkeysvalueof{/pgfplots/parent axis height},
%		plot graphics/node/.style={scale=1.33,anchor=south west,inner sep=0pt,}, %%% scale colorbar fill %%%
            ytick={0.00,0.20,0.40,0.60,0.80,1.00},
            yticklabels={0.00,0.20,0.40,0.60,0.80,1.00},
            yticklabel={\pgfmathprintnumber\tick},
            yticklabel style={%font=\footnotesize,
            		scale=1.25,
            		/pgf/number format/fixed,
			/pgf/number format/fixed zerofill,
			/pgf/number format/precision=2}
        },
%        point meta min=0,
%        point meta max=200,
        point meta min=0.0,
        point meta max=1.0,
        nodes near coords={\pgfmathprintnumber\pgfplotspointmeta},
        % ---------------------------------------------------------------------
        % show `nodes near coords' but adapt the style so that values
        % above a threshold get another style
        % (adapted from <http://tex.stackexchange.com/a/141006/95441>)
        % #1: the THRESHOLD after which we switch to a special display.
        nodes near coords black white/.style={
            % define the style of the nodes with "small" values
            small value/.style={
                yshift=-7pt,
%                text=white,
                text=black,
                /pgf/number format/fixed,
%                /pgf/number format/precision=0,
                /pgf/number format/precision=2,
                /pgf/number format/zerofill=true,
                scale=0.95,
%                /pgf/number format/precision=0
            },
            % define the style of the nodes with "large" values
            large value/.style={
                yshift=-7pt,
%                text=black,
                text=white,
                /pgf/number format/fixed,
%                /pgf/number format/precision=2,
                /pgf/number format/precision=2,
                /pgf/number format/zerofill=true,
                scale=0.95,
%                /pgf/number format/precision=0
            },
            every node near coord/.style={
                check for zero/.code={
                    \pgfmathfloatifflags{\pgfplotspointmeta}{0}{
                        % If meta=0, make the node a coordinate
                        % (which doesn't have text)
                        \pgfkeys{/tikz/coordinate}
                    }{
                        \begingroup
                        % this group is merely to switch to FPU locally.
                        % Might be unnecessary, but who knows.
                        \pgfkeys{/pgf/fpu}
                        \pgfmathparse{\pgfplotspointmeta<#1}
                        \global\let\result=\pgfmathresult
                        \endgroup
                        %
                        % simplifies debugging:
                        %\show\result
                        %
                        \pgfmathfloatcreate{1}{1.0}{0}
                        \let\ONE=\pgfmathresult
                        \ifx\result\ONE
                            % AH: our condition 'y < #1' is met.
                            \pgfkeysalso{/pgfplots/small value}
                        \else
                            % ok, proceed as usual.
                            \pgfkeysalso{/pgfplots/large value}
                        \fi
                    }
                },
                check for zero,
            },
        },
        % asign a value to the new style which is the threshold at which
        % the two style `small value' or `large value' are used
%        nodes near coords black white=100,
        nodes near coords black white=0.5,
        % -----------------------------------------------------------------
    ]
\addplot[
  matrix plot,
  mesh/cols=17,
  point meta=explicit,
  draw=gray
] table [x=x, y=y, meta=C] {
x y C
0 0 0.51
1 0 0.04
2 0 0.00
3 0 0.03
4 0 0.00
5 0 0.00
6 0 0.03
7 0 0.00
8 0 0.00
9 0 0.06
10 0 0.06
11 0 0.05
12 0 0.06
13 0 0.01
14 0 0.00
15 0 0.14
16 0 0.01
0 1 0.02
1 1 0.62
2 1 0.00
3 1 0.05
4 1 0.03
5 1 0.01
6 1 0.02
7 1 0.05
8 1 0.00
9 1 0.04
10 1 0.01
11 1 0.05
12 1 0.03
13 1 0.02
14 1 0.02
15 1 0.03
16 1 0.00
0 2 0.01
1 2 0.01
2 2 0.83
3 2 0.00
4 2 0.00
5 2 0.00
6 2 0.01
7 2 0.01
8 2 0.04
9 2 0.00
10 2 0.01
11 2 0.01
12 2 0.00
13 2 0.03
14 2 0.02
15 2 0.00
16 2 0.02
0 3 0.09
1 3 0.04
2 3 0.01
3 3 0.66
4 3 0.00
5 3 0.02
6 3 0.00
7 3 0.01
8 3 0.00
9 3 0.05
10 3 0.04
11 3 0.00
12 3 0.03
13 3 0.01
14 3 0.00
15 3 0.04
16 3 0.00
0 4 0.06
1 4 0.01
2 4 0.00
3 4 0.03
4 4 0.68
5 4 0.02
6 4 0.06
7 4 0.00
8 4 0.00
9 4 0.02
10 4 0.01
11 4 0.06
12 4 0.00
13 4 0.03
14 4 0.01
15 4 0.00
16 4 0.01
0 5 0.01
1 5 0.01
2 5 0.00
3 5 0.00
4 5 0.01
5 5 0.93
6 5 0.00
7 5 0.00
8 5 0.00
9 5 0.01
10 5 0.00
11 5 0.00
12 5 0.00
13 5 0.00
14 5 0.01
15 5 0.02
16 5 0.00
0 6 0.02
1 6 0.02
2 6 0.01
3 6 0.03
4 6 0.04
5 6 0.00
6 6 0.67
7 6 0.00
8 6 0.01
9 6 0.01
10 6 0.01
11 6 0.04
12 6 0.01
13 6 0.10
14 6 0.02
15 6 0.01
16 6 0.00
0 7 0.00
1 7 0.07
2 7 0.00
3 7 0.00
4 7 0.00
5 7 0.00
6 7 0.00
7 7 0.82
8 7 0.00
9 7 0.01
10 7 0.00
11 7 0.00
12 7 0.01
13 7 0.00
14 7 0.09
15 7 0.00
16 7 0.00
0 8 0.00
1 8 0.00
2 8 0.00
3 8 0.01
4 8 0.00
5 8 0.00
6 8 0.02
7 8 0.00
8 8 0.78
9 8 0.00
10 8 0.00
11 8 0.01
12 8 0.00
13 8 0.07
14 8 0.01
15 8 0.00
16 8 0.10
0 9 0.11
1 9 0.05
2 9 0.00
3 9 0.07
4 9 0.02
5 9 0.00
6 9 0.02
7 9 0.01
8 9 0.00
9 9 0.44
10 9 0.05
11 9 0.04
12 9 0.04
13 9 0.00
14 9 0.00
15 9 0.15
16 9 0.00
0 10 0.03
1 10 0.05
2 10 0.02
3 10 0.02
4 10 0.00
5 10 0.00
6 10 0.08
7 10 0.01
8 10 0.00
9 10 0.05
10 10 0.64
11 10 0.03
12 10 0.00
13 10 0.00
14 10 0.04
15 10 0.03
16 10 0.00
0 11 0.04
1 11 0.01
2 11 0.00
3 11 0.02
4 11 0.04
5 11 0.00
6 11 0.09
7 11 0.01
8 11 0.00
9 11 0.00
10 11 0.03
11 11 0.72
12 11 0.03
13 11 0.00
14 11 0.00
15 11 0.01
16 11 0.00
0 12 0.05
1 12 0.03
2 12 0.00
3 12 0.02
4 12 0.02
5 12 0.00
6 12 0.00
7 12 0.01
8 12 0.00
9 12 0.01
10 12 0.01
11 12 0.01
12 12 0.81
13 12 0.00
14 12 0.01
15 12 0.02
16 12 0.00
0 13 0.01
1 13 0.00
2 13 0.04
3 13 0.00
4 13 0.01
5 13 0.02
6 13 0.09
7 13 0.00
8 13 0.07
9 13 0.04
10 13 0.01
11 13 0.01
12 13 0.00
13 13 0.68
14 13 0.01
15 13 0.00
16 13 0.01
0 14 0.01
1 14 0.03
2 14 0.01
3 14 0.01
4 14 0.00
5 14 0.00
6 14 0.02
7 14 0.04
8 14 0.02
9 14 0.04
10 14 0.00
11 14 0.00
12 14 0.00
13 14 0.09
14 14 0.67
15 14 0.02
16 14 0.04
0 15 0.09
1 15 0.05
2 15 0.00
3 15 0.07
4 15 0.00
5 15 0.00
6 15 0.02
7 15 0.00
8 15 0.00
9 15 0.14
10 15 0.01
11 15 0.00
12 15 0.03
13 15 0.00
14 15 0.00
15 15 0.59
16 15 0.00
0 16 0.00
1 16 0.01
2 16 0.01
3 16 0.01
4 16 0.00
5 16 0.00
6 16 0.02
7 16 0.01
8 16 0.17
9 16 0.00
10 16 0.00
11 16 0.00
12 16 0.00
13 16 0.06
14 16 0.02
15 16 0.00
16 16 0.69
};
\end{axis}
%\draw[black,thick] (2.625,5.8) circle(0.5);
%\draw[red,dashed,thick] (3.675,5.8) circle(0.5);
\end{tikzpicture}
\caption{KNN baseline confusion matrix (accuracy~0.6906)}\label{fig:conf_KB}
\end{figure}
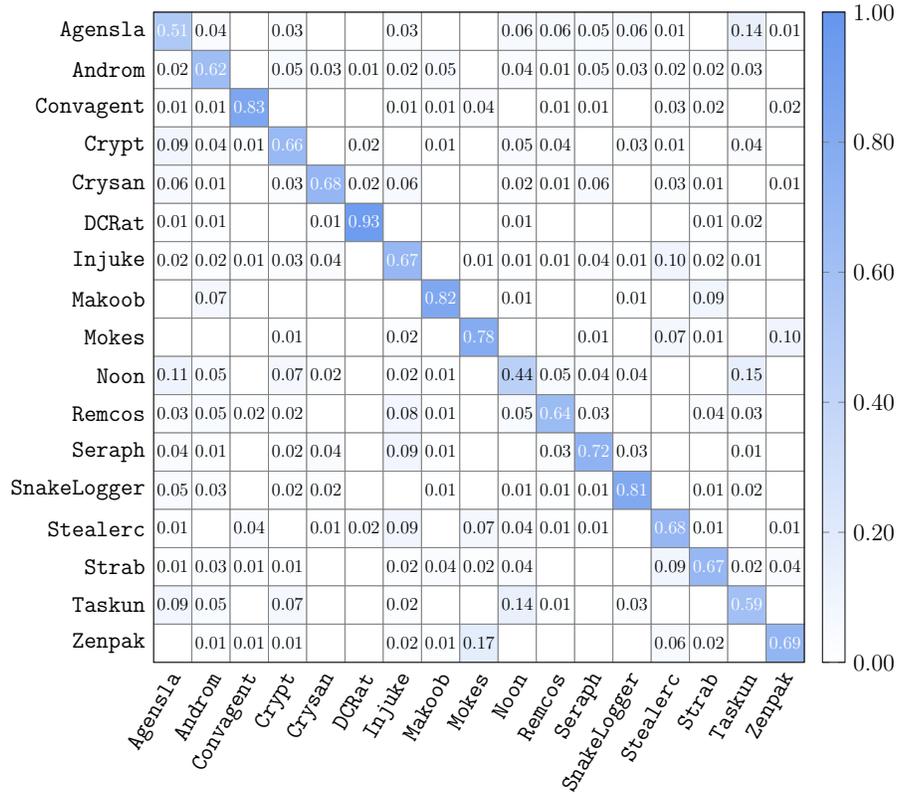

\begin{figure}[!htb]
\centering
%\begin{tikzpicture}[scale=0.8,every node/.style={scale=0.8}]
\begin{tikzpicture}[scale=0.60]
    \begin{axis}[%colorbar/width=2.5mm,
        width=16cm,
        height=16cm,
%        xlabel={\LARGE Predicted class},
%        ylabel={\LARGE Actual class},
%        colormap={blackwhite}{gray(0cm)=(1); gray(1cm)=(0.5)},
%	colormap={bluewhite}{color=(white) color=(blue)},
%	colormap={bluewhite}{color=(white) rgb255=(0,191,255)},
	colormap={bluewhite}{color=(white) rgb255=(100,149,237)},
        xticklabels={
        Agensla,
        Androm,
        Convagent,
        Crypt,
        Crysan,
        DCRat,
        Injuke,
        Makoob,
        Mokes,
        Noon,
        Remcos,
        Seraph,
        SnakeLogger,
        Stealerc,
        Strab,
        Taskun,
        Zenpak
        },
        xtick={0,...,16},
        xtick style={draw=none},
	xticklabel style={scale=1.25,anchor=east,rotate=60,yshift=-5pt,font=\tt},
        yticklabels={
        Agensla,
        Androm,
        Convagent,
        Crypt,
        Crysan,
        DCRat,
        Injuke,
        Makoob,
        Mokes,
        Noon,
        Remcos,
        Seraph,
        SnakeLogger,
        Stealerc,
        Strab,
        Taskun,
        Zenpak
        },
        ytick={0,...,16},
        ytick style={draw=none},
        enlargelimits=false,
        yticklabel style={scale=1.25,font=\tt},
        colorbar,
        colorbar style={
%     	  	width=0.05*\pgfkeysvalueof{/pgfplots/parent axis width},%%% added this
%     	  	height=0.5*\pgfkeysvalueof{/pgfplots/parent axis height},
%		plot graphics/node/.style={scale=1.33,anchor=south west,inner sep=0pt,}, %%% scale colorbar fill %%%
            ytick={0.00,0.20,0.40,0.60,0.80,1.00},
            yticklabels={0.00,0.20,0.40,0.60,0.80,1.00},
            yticklabel={\pgfmathprintnumber\tick},
            yticklabel style={%font=\footnotesize,
            		scale=1.25,
            		/pgf/number format/fixed,
			/pgf/number format/fixed zerofill,
			/pgf/number format/precision=2}
        },
%        point meta min=0,
%        point meta max=200,
        point meta min=0.0,
        point meta max=1.0,
        nodes near coords={\pgfmathprintnumber\pgfplotspointmeta},
        % ---------------------------------------------------------------------
        % show `nodes near coords' but adapt the style so that values
        % above a threshold get another style
        % (adapted from <http://tex.stackexchange.com/a/141006/95441>)
        % #1: the THRESHOLD after which we switch to a special display.
        nodes near coords black white/.style={
            % define the style of the nodes with "small" values
            small value/.style={
                yshift=-7pt,
%                text=white,
                text=black,
                /pgf/number format/fixed,
%                /pgf/number format/precision=0,
                /pgf/number format/precision=2,
                /pgf/number format/zerofill=true,
                scale=0.95,
%                /pgf/number format/precision=0
            },
            % define the style of the nodes with "large" values
            large value/.style={
                yshift=-7pt,
%                text=black,
                text=white,
                /pgf/number format/fixed,
%                /pgf/number format/precision=2,
                /pgf/number format/precision=2,
                /pgf/number format/zerofill=true,
                scale=0.95,
%                /pgf/number format/precision=0
            },
            every node near coord/.style={
                check for zero/.code={
                    \pgfmathfloatifflags{\pgfplotspointmeta}{0}{
                        % If meta=0, make the node a coordinate
                        % (which doesn't have text)
                        \pgfkeys{/tikz/coordinate}
                    }{
                        \begingroup
                        % this group is merely to switch to FPU locally.
                        % Might be unnecessary, but who knows.
                        \pgfkeys{/pgf/fpu}
                        \pgfmathparse{\pgfplotspointmeta<#1}
                        \global\let\result=\pgfmathresult
                        \endgroup
                        %
                        % simplifies debugging:
                        %\show\result
                        %
                        \pgfmathfloatcreate{1}{1.0}{0}
                        \let\ONE=\pgfmathresult
                        \ifx\result\ONE
                            % AH: our condition 'y < #1' is met.
                            \pgfkeysalso{/pgfplots/small value}
                        \else
                            % ok, proceed as usual.
                            \pgfkeysalso{/pgfplots/large value}
                        \fi
                    }
                },
                check for zero,
            },
        },
        % asign a value to the new style which is the threshold at which
        % the two style `small value' or `large value' are used
%        nodes near coords black white=100,
        nodes near coords black white=0.5,
        % -----------------------------------------------------------------
    ]
\addplot[
  matrix plot,
  mesh/cols=17,
  point meta=explicit,
  draw=gray
] table [x=x, y=y, meta=C] {
x y C
0 0 0.65
1 0 0.02
2 0 0.00
3 0 0.02
4 0 0.01
5 0 0.00
6 0 0.01
7 0 0.01
8 0 0.00
9 0 0.08
10 0 0.02
11 0 0.06
12 0 0.00
13 0 0.01
14 0 0.00
15 0 0.11
16 0 0.00
0 1 0.01
1 1 0.67
2 1 0.04
3 1 0.02
4 1 0.00
5 1 0.00
6 1 0.05
7 1 0.03
8 1 0.00
9 1 0.02
10 1 0.01
11 1 0.06
12 1 0.00
13 1 0.04
14 1 0.03
15 1 0.02
16 1 0.00
0 2 0.00
1 2 0.00
2 2 0.77
3 2 0.00
4 2 0.00
5 2 0.00
6 2 0.08
7 2 0.00
8 2 0.01
9 2 0.00
10 2 0.00
11 2 0.01
12 2 0.01
13 2 0.08
14 2 0.01
15 2 0.00
16 2 0.03
0 3 0.01
1 3 0.02
2 3 0.00
3 3 0.71
4 3 0.01
5 3 0.00
6 3 0.06
7 3 0.00
8 3 0.00
9 3 0.01
10 3 0.02
11 3 0.10
12 3 0.00
13 3 0.04
14 3 0.00
15 3 0.02
16 3 0.00
0 4 0.01
1 4 0.00
2 4 0.00
3 4 0.02
4 4 0.77
5 4 0.00
6 4 0.04
7 4 0.00
8 4 0.00
9 4 0.00
10 4 0.03
11 4 0.07
12 4 0.00
13 4 0.02
14 4 0.02
15 4 0.02
16 4 0.00
0 5 0.01
1 5 0.00
2 5 0.00
3 5 0.00
4 5 0.00
5 5 0.95
6 5 0.01
7 5 0.00
8 5 0.00
9 5 0.00
10 5 0.00
11 5 0.01
12 5 0.02
13 5 0.00
14 5 0.00
15 5 0.00
16 5 0.00
0 6 0.04
1 6 0.00
2 6 0.04
3 6 0.01
4 6 0.00
5 6 0.00
6 6 0.58
7 6 0.00
8 6 0.01
9 6 0.02
10 6 0.01
11 6 0.13
12 6 0.00
13 6 0.13
14 6 0.02
15 6 0.00
16 6 0.01
0 7 0.00
1 7 0.01
2 7 0.00
3 7 0.00
4 7 0.00
5 7 0.00
6 7 0.00
7 7 0.98
8 7 0.00
9 7 0.00
10 7 0.00
11 7 0.00
12 7 0.00
13 7 0.00
14 7 0.01
15 7 0.00
16 7 0.00
0 8 0.00
1 8 0.00
2 8 0.02
3 8 0.00
4 8 0.00
5 8 0.00
6 8 0.02
7 8 0.00
8 8 0.73
9 8 0.00
10 8 0.00
11 8 0.00
12 8 0.00
13 8 0.08
14 8 0.01
15 8 0.00
16 8 0.14
0 9 0.09
1 9 0.05
2 9 0.00
3 9 0.02
4 9 0.02
5 9 0.00
6 9 0.03
7 9 0.00
8 9 0.00
9 9 0.61
10 9 0.01
11 9 0.06
12 9 0.02
13 9 0.01
14 9 0.01
15 9 0.07
16 9 0.00
0 10 0.03
1 10 0.00
2 10 0.01
3 10 0.04
4 10 0.02
5 10 0.00
6 10 0.02
7 10 0.01
8 10 0.00
9 10 0.01
10 10 0.72
11 10 0.10
12 10 0.02
13 10 0.02
14 10 0.00
15 10 0.00
16 10 0.00
0 11 0.09
1 11 0.00
2 11 0.00
3 11 0.04
4 11 0.04
5 11 0.00
6 11 0.03
7 11 0.00
8 11 0.00
9 11 0.00
10 11 0.01
11 11 0.73
12 11 0.00
13 11 0.04
14 11 0.00
15 11 0.02
16 11 0.00
0 12 0.04
1 12 0.00
2 12 0.00
3 12 0.05
4 12 0.01
5 12 0.00
6 12 0.00
7 12 0.00
8 12 0.01
9 12 0.03
10 12 0.00
11 12 0.05
12 12 0.76
13 12 0.01
14 12 0.00
15 12 0.04
16 12 0.00
0 13 0.01
1 13 0.02
2 13 0.02
3 13 0.03
4 13 0.00
5 13 0.00
6 13 0.04
7 13 0.00
8 13 0.10
9 13 0.00
10 13 0.00
11 13 0.03
12 13 0.00
13 13 0.63
14 13 0.10
15 13 0.00
16 13 0.02
0 14 0.01
1 14 0.00
2 14 0.02
3 14 0.00
4 14 0.00
5 14 0.00
6 14 0.01
7 14 0.02
8 14 0.02
9 14 0.00
10 14 0.02
11 14 0.00
12 14 0.00
13 14 0.05
14 14 0.84
15 14 0.00
16 14 0.01
0 15 0.14
1 15 0.00
2 15 0.00
3 15 0.01
4 15 0.00
5 15 0.00
6 15 0.00
7 15 0.00
8 15 0.00
9 15 0.05
10 15 0.00
11 15 0.05
12 15 0.00
13 15 0.01
14 15 0.00
15 15 0.74
16 15 0.00
0 16 0.00
1 16 0.00
2 16 0.03
3 16 0.01
4 16 0.00
5 16 0.00
6 16 0.01
7 16 0.01
8 16 0.13
9 16 0.01
10 16 0.00
11 16 0.00
12 16 0.00
13 16 0.06
14 16 0.06
15 16 0.00
16 16 0.68
};
\end{axis}
%\draw[black,thick] (2.625,5.8) circle(0.5);
%\draw[red,dashed,thick] (3.675,5.8) circle(0.5);
\end{tikzpicture}
\caption{Grayscale XGBoost HOG confusion matrix (accuracy~0.7512)}\label{fig:conf_XHG}
\end{figure}
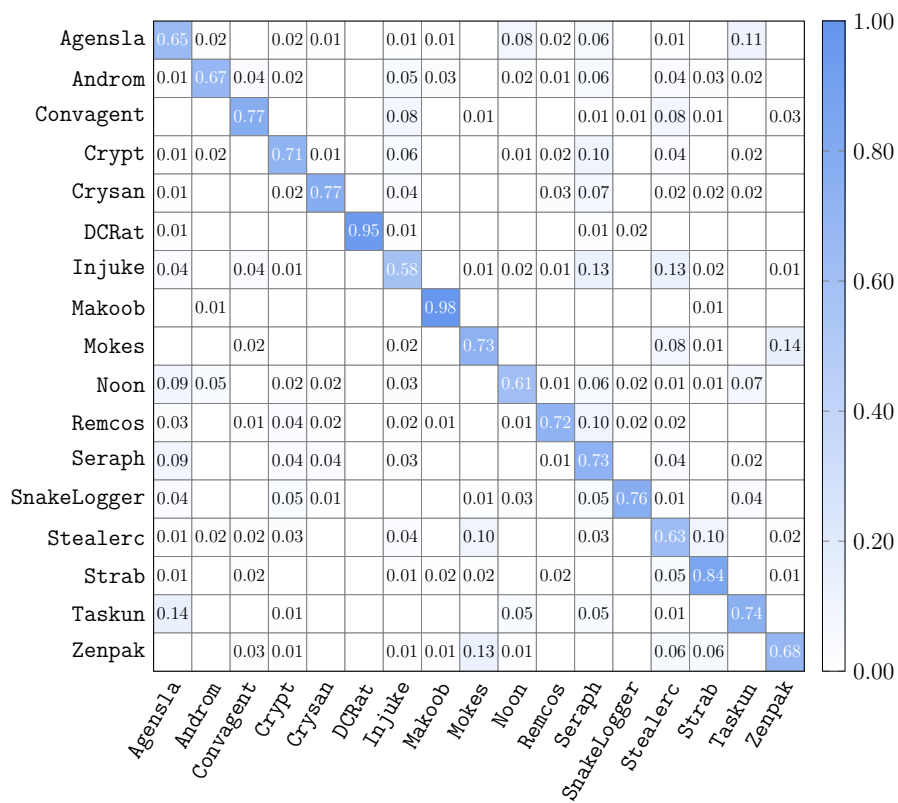

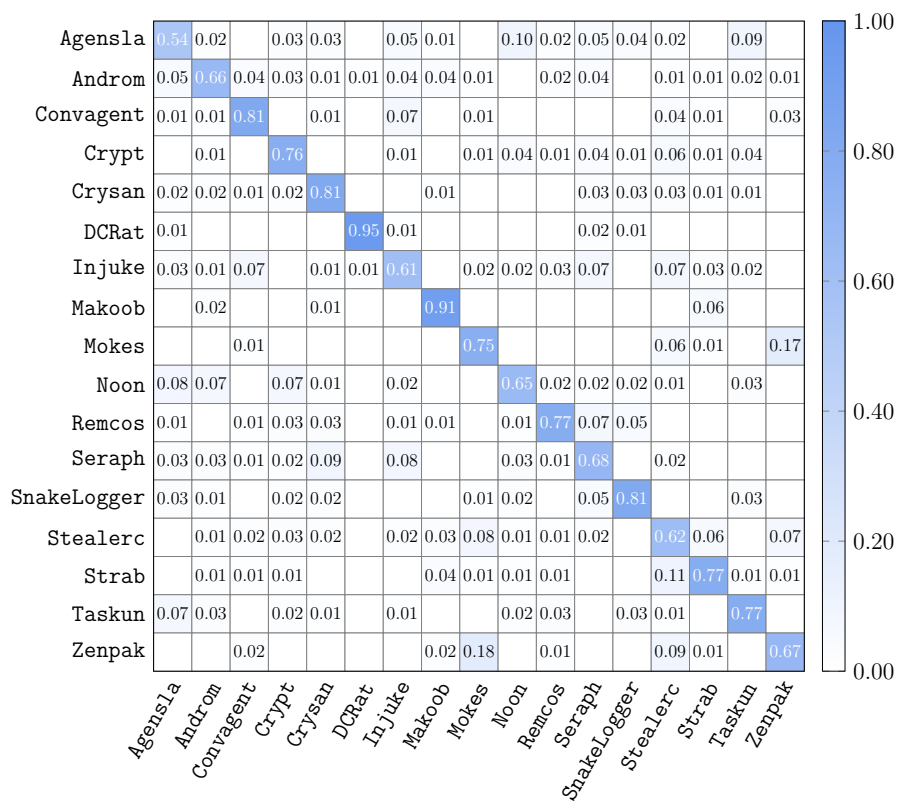
\begin{figure}[!htb]
\centering
%\begin{tikzpicture}[scale=0.8,every node/.style={scale=0.8}]
\begin{tikzpicture}[scale=0.60]
    \begin{axis}[%colorbar/width=2.5mm,
        width=16cm,
        height=16cm,
%        xlabel={\LARGE Predicted class},
%        ylabel={\LARGE Actual class},
%        colormap={blackwhite}{gray(0cm)=(1); gray(1cm)=(0.5)},
%	colormap={bluewhite}{color=(white) color=(blue)},
%	colormap={bluewhite}{color=(white) rgb255=(0,191,255)},
	colormap={bluewhite}{color=(white) rgb255=(100,149,237)},
        xticklabels={
        Agensla,
        Androm,
        Convagent,
        Crypt,
        Crysan,
        DCRat,
        Injuke,
        Makoob,
        Mokes,
        Noon,
        Remcos,
        Seraph,
        SnakeLogger,
        Stealerc,
        Strab,
        Taskun,
        Zenpak
        },
        xtick={0,...,16},
        xtick style={draw=none},
	xticklabel style={scale=1.25,anchor=east,rotate=60,yshift=-5pt,font=\tt},
        yticklabels={
        Agensla,
        Androm,
        Convagent,
        Crypt,
        Crysan,
        DCRat,
        Injuke,
        Makoob,
        Mokes,
        Noon,
        Remcos,
        Seraph,
        SnakeLogger,
        Stealerc,
        Strab,
        Taskun,
        Zenpak
        },
        ytick={0,...,16},
        ytick style={draw=none},
        enlargelimits=false,
        yticklabel style={scale=1.25,font=\tt},
        colorbar,
        colorbar style={
%     	  	width=0.05*\pgfkeysvalueof{/pgfplots/parent axis width},%%% added this
%     	  	height=0.5*\pgfkeysvalueof{/pgfplots/parent axis height},
%		plot graphics/node/.style={scale=1.33,anchor=south west,inner sep=0pt,}, %%% scale colorbar fill %%%
            ytick={0.00,0.20,0.40,0.60,0.80,1.00},
            yticklabels={0.00,0.20,0.40,0.60,0.80,1.00},
            yticklabel={\pgfmathprintnumber\tick},
            yticklabel style={%font=\footnotesize,
            		scale=1.25,
            		/pgf/number format/fixed,
			/pgf/number format/fixed zerofill,
			/pgf/number format/precision=2}
        },
%        point meta min=0,
%        point meta max=200,
        point meta min=0.0,
        point meta max=1.0,
        nodes near coords={\pgfmathprintnumber\pgfplotspointmeta},
        % ---------------------------------------------------------------------
        % show `nodes near coords' but adapt the style so that values
        % above a threshold get another style
        % (adapted from <http://tex.stackexchange.com/a/141006/95441>)
        % #1: the THRESHOLD after which we switch to a special display.
        nodes near coords black white/.style={
            % define the style of the nodes with "small" values
            small value/.style={
                yshift=-7pt,
%                text=white,
                text=black,
                /pgf/number format/fixed,
%                /pgf/number format/precision=0,
                /pgf/number format/precision=2,
                /pgf/number format/zerofill=true,
                scale=0.95,
%                /pgf/number format/precision=0
            },
            % define the style of the nodes with "large" values
            large value/.style={
                yshift=-7pt,
%                text=black,
                text=white,
                /pgf/number format/fixed,
%                /pgf/number format/precision=2,
                /pgf/number format/precision=2,
                /pgf/number format/zerofill=true,
                scale=0.95,
%                /pgf/number format/precision=0
            },
            every node near coord/.style={
                check for zero/.code={
                    \pgfmathfloatifflags{\pgfplotspointmeta}{0}{
                        % If meta=0, make the node a coordinate
                        % (which doesn't have text)
                        \pgfkeys{/tikz/coordinate}
                    }{
                        \begingroup
                        % this group is merely to switch to FPU locally.
                        % Might be unnecessary, but who knows.
                        \pgfkeys{/pgf/fpu}
                        \pgfmathparse{\pgfplotspointmeta<#1}
                        \global\let\result=\pgfmathresult
                        \endgroup
                        %
                        % simplifies debugging:
                        %\show\result
                        %
                        \pgfmathfloatcreate{1}{1.0}{0}
                        \let\ONE=\pgfmathresult
                        \ifx\result\ONE
                            % AH: our condition 'y < #1' is met.
                            \pgfkeysalso{/pgfplots/small value}
                        \else
                            % ok, proceed as usual.
                            \pgfkeysalso{/pgfplots/large value}
                        \fi
                    }
                },
                check for zero,
            },
        },
        % asign a value to the new style which is the threshold at which
        % the two style `small value' or `large value' are used
%        nodes near coords black white=100,
        nodes near coords black white=0.5,
        % -----------------------------------------------------------------
    ]
\addplot[
  matrix plot,
  mesh/cols=17,
  point meta=explicit,
  draw=gray
] table [x=x, y=y, meta=C] {
x y C
0 0 0.54
1 0 0.02
2 0 0.00
3 0 0.03
4 0 0.03
5 0 0.00
6 0 0.05
7 0 0.01
8 0 0.00
9 0 0.10
10 0 0.02
11 0 0.05
12 0 0.04
13 0 0.02
14 0 0.00
15 0 0.09
16 0 0.00
0 1 0.05
1 1 0.66
2 1 0.04
3 1 0.03
4 1 0.01
5 1 0.01
6 1 0.04
7 1 0.04
8 1 0.01
9 1 0.00
10 1 0.02
11 1 0.04
12 1 0.00
13 1 0.01
14 1 0.01
15 1 0.02
16 1 0.01
0 2 0.01
1 2 0.01
2 2 0.81
3 2 0.00
4 2 0.01
5 2 0.00
6 2 0.07
7 2 0.00
8 2 0.01
9 2 0.00
10 2 0.00
11 2 0.00
12 2 0.00
13 2 0.04
14 2 0.01
15 2 0.00
16 2 0.03
0 3 0.00
1 3 0.01
2 3 0.00
3 3 0.76
4 3 0.00
5 3 0.00
6 3 0.01
7 3 0.00
8 3 0.01
9 3 0.04
10 3 0.01
11 3 0.04
12 3 0.01
13 3 0.06
14 3 0.01
15 3 0.04
16 3 0.00
0 4 0.02
1 4 0.02
2 4 0.01
3 4 0.02
4 4 0.81
5 4 0.00
6 4 0.00
7 4 0.01
8 4 0.00
9 4 0.00
10 4 0.00
11 4 0.03
12 4 0.03
13 4 0.03
14 4 0.01
15 4 0.01
16 4 0.00
0 5 0.01
1 5 0.00
2 5 0.00
3 5 0.00
4 5 0.00
5 5 0.95
6 5 0.01
7 5 0.00
8 5 0.00
9 5 0.00
10 5 0.00
11 5 0.02
12 5 0.01
13 5 0.00
14 5 0.00
15 5 0.00
16 5 0.00
0 6 0.03
1 6 0.01
2 6 0.07
3 6 0.00
4 6 0.01
5 6 0.01
6 6 0.61
7 6 0.00
8 6 0.02
9 6 0.02
10 6 0.03
11 6 0.07
12 6 0.00
13 6 0.07
14 6 0.03
15 6 0.02
16 6 0.00
0 7 0.00
1 7 0.02
2 7 0.00
3 7 0.00
4 7 0.01
5 7 0.00
6 7 0.00
7 7 0.91
8 7 0.00
9 7 0.00
10 7 0.00
11 7 0.00
12 7 0.00
13 7 0.00
14 7 0.06
15 7 0.00
16 7 0.00
0 8 0.00
1 8 0.00
2 8 0.01
3 8 0.00
4 8 0.00
5 8 0.00
6 8 0.00
7 8 0.00
8 8 0.75
9 8 0.00
10 8 0.00
11 8 0.00
12 8 0.00
13 8 0.06
14 8 0.01
15 8 0.00
16 8 0.17
0 9 0.08
1 9 0.07
2 9 0.00
3 9 0.07
4 9 0.01
5 9 0.00
6 9 0.02
7 9 0.00
8 9 0.00
9 9 0.65
10 9 0.02
11 9 0.02
12 9 0.02
13 9 0.01
14 9 0.00
15 9 0.03
16 9 0.00
0 10 0.01
1 10 0.00
2 10 0.01
3 10 0.03
4 10 0.03
5 10 0.00
6 10 0.01
7 10 0.01
8 10 0.00
9 10 0.01
10 10 0.77
11 10 0.07
12 10 0.05
13 10 0.00
14 10 0.00
15 10 0.00
16 10 0.00
0 11 0.03
1 11 0.03
2 11 0.01
3 11 0.02
4 11 0.09
5 11 0.00
6 11 0.08
7 11 0.00
8 11 0.00
9 11 0.03
10 11 0.01
11 11 0.68
12 11 0.00
13 11 0.02
14 11 0.00
15 11 0.00
16 11 0.00
0 12 0.03
1 12 0.01
2 12 0.00
3 12 0.02
4 12 0.02
5 12 0.00
6 12 0.00
7 12 0.00
8 12 0.01
9 12 0.02
10 12 0.00
11 12 0.05
12 12 0.81
13 12 0.00
14 12 0.00
15 12 0.03
16 12 0.00
0 13 0.00
1 13 0.01
2 13 0.02
3 13 0.03
4 13 0.02
5 13 0.00
6 13 0.02
7 13 0.03
8 13 0.08
9 13 0.01
10 13 0.01
11 13 0.02
12 13 0.00
13 13 0.62
14 13 0.06
15 13 0.00
16 13 0.07
0 14 0.00
1 14 0.01
2 14 0.01
3 14 0.01
4 14 0.00
5 14 0.00
6 14 0.00
7 14 0.04
8 14 0.01
9 14 0.01
10 14 0.01
11 14 0.00
12 14 0.00
13 14 0.11
14 14 0.77
15 14 0.01
16 14 0.01
0 15 0.07
1 15 0.03
2 15 0.00
3 15 0.02
4 15 0.01
5 15 0.00
6 15 0.01
7 15 0.00
8 15 0.00
9 15 0.02
10 15 0.03
11 15 0.00
12 15 0.03
13 15 0.01
14 15 0.00
15 15 0.77
16 15 0.00
0 16 0.00
1 16 0.00
2 16 0.02
3 16 0.00
4 16 0.00
5 16 0.00
6 16 0.00
7 16 0.02
8 16 0.18
9 16 0.00
10 16 0.01
11 16 0.00
12 16 0.00
13 16 0.09
14 16 0.01
15 16 0.00
16 16 0.67
};
\end{axis}
%\draw[black,thick] (2.625,5.8) circle(0.5);
%\draw[red,dashed,thick] (3.675,5.8) circle(0.5);
\end{tikzpicture}
\caption{Grayscale XGBoost Haralick confusion matrix (accuracy~0.7229)}\label{fig:conf_XHarG}
\end{figure}

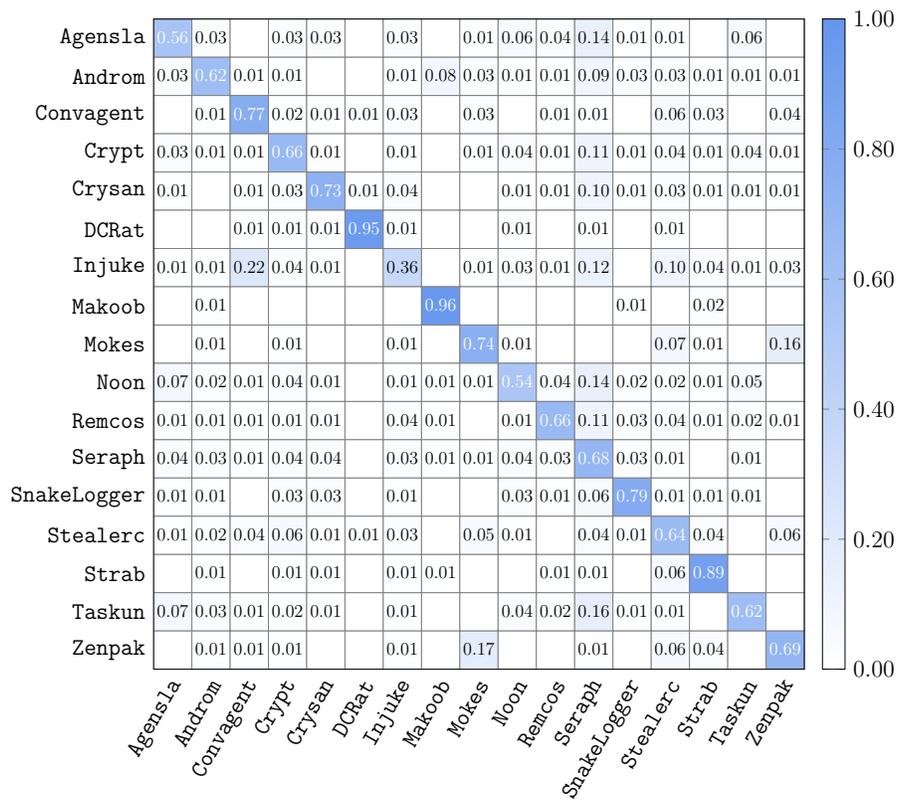
\begin{figure}[!htb]
\centering
%\begin{tikzpicture}[scale=0.8,every node/.style={scale=0.8}]
\begin{tikzpicture}[scale=0.60]
    \begin{axis}[%colorbar/width=2.5mm,
        width=16cm,
        height=16cm,
%        xlabel={\LARGE Predicted class},
%        ylabel={\LARGE Actual class},
%        colormap={blackwhite}{gray(0cm)=(1); gray(1cm)=(0.5)},
%	colormap={bluewhite}{color=(white) color=(blue)},
%	colormap={bluewhite}{color=(white) rgb255=(0,191,255)},
	colormap={bluewhite}{color=(white) rgb255=(100,149,237)},
        xticklabels={
        Agensla,
        Androm,
        Convagent,
        Crypt,
        Crysan,
        DCRat,
        Injuke,
        Makoob,
        Mokes,
        Noon,
        Remcos,
        Seraph,
        SnakeLogger,
        Stealerc,
        Strab,
        Taskun,
        Zenpak
        },
        xtick={0,...,16},
        xtick style={draw=none},
	xticklabel style={scale=1.25,anchor=east,rotate=60,yshift=-5pt,font=\tt},
        yticklabels={
        Agensla,
        Androm,
        Convagent,
        Crypt,
        Crysan,
        DCRat,
        Injuke,
        Makoob,
        Mokes,
        Noon,
        Remcos,
        Seraph,
        SnakeLogger,
        Stealerc,
        Strab,
        Taskun,
        Zenpak
        },
        ytick={0,...,16},
        ytick style={draw=none},
        enlargelimits=false,
        yticklabel style={scale=1.25,font=\tt},
        colorbar,
        colorbar style={
%     	  	width=0.05*\pgfkeysvalueof{/pgfplots/parent axis width},%%% added this
%     	  	height=0.5*\pgfkeysvalueof{/pgfplots/parent axis height},
%		plot graphics/node/.style={scale=1.33,anchor=south west,inner sep=0pt,}, %%% scale colorbar fill %%%
            ytick={0.00,0.20,0.40,0.60,0.80,1.00},
            yticklabels={0.00,0.20,0.40,0.60,0.80,1.00},
            yticklabel={\pgfmathprintnumber\tick},
            yticklabel style={%font=\footnotesize,
            		scale=1.25,
            		/pgf/number format/fixed,
			/pgf/number format/fixed zerofill,
			/pgf/number format/precision=2}
        },
%        point meta min=0,
%        point meta max=200,
        point meta min=0.0,
        point meta max=1.0,
        nodes near coords={\pgfmathprintnumber\pgfplotspointmeta},
        % ---------------------------------------------------------------------
        % show `nodes near coords' but adapt the style so that values
        % above a threshold get another style
        % (adapted from <http://tex.stackexchange.com/a/141006/95441>)
        % #1: the THRESHOLD after which we switch to a special display.
        nodes near coords black white/.style={
            % define the style of the nodes with "small" values
            small value/.style={
                yshift=-7pt,
%                text=white,
                text=black,
                /pgf/number format/fixed,
%                /pgf/number format/precision=0,
                /pgf/number format/precision=2,
                /pgf/number format/zerofill=true,
                scale=0.95,
%                /pgf/number format/precision=0
            },
            % define the style of the nodes with "large" values
            large value/.style={
                yshift=-7pt,
%                text=black,
                text=white,
                /pgf/number format/fixed,
%                /pgf/number format/precision=2,
                /pgf/number format/precision=2,
                /pgf/number format/zerofill=true,
                scale=0.95,
%                /pgf/number format/precision=0
            },
            every node near coord/.style={
                check for zero/.code={
                    \pgfmathfloatifflags{\pgfplotspointmeta}{0}{
                        % If meta=0, make the node a coordinate
                        % (which doesn't have text)
                        \pgfkeys{/tikz/coordinate}
                    }{
                        \begingroup
                        % this group is merely to switch to FPU locally.
                        % Might be unnecessary, but who knows.
                        \pgfkeys{/pgf/fpu}
                        \pgfmathparse{\pgfplotspointmeta<#1}
                        \global\let\result=\pgfmathresult
                        \endgroup
                        %
                        % simplifies debugging:
                        %\show\result
                        %
                        \pgfmathfloatcreate{1}{1.0}{0}
                        \let\ONE=\pgfmathresult
                        \ifx\result\ONE
                            % AH: our condition 'y < #1' is met.
                            \pgfkeysalso{/pgfplots/small value}
                        \else
                            % ok, proceed as usual.
                            \pgfkeysalso{/pgfplots/large value}
                        \fi
                    }
                },
                check for zero,
            },
        },
        % asign a value to the new style which is the threshold at which
        % the two style `small value' or `large value' are used
%        nodes near coords black white=100,
        nodes near coords black white=0.5,
        % -----------------------------------------------------------------
    ]
\addplot[
  matrix plot,
  mesh/cols=17,
  point meta=explicit,
  draw=gray
] table [x=x, y=y, meta=C] {
x y C
0 0 0.56
1 0 0.03
2 0 0.00
3 0 0.03
4 0 0.03
5 0 0.00
6 0 0.03
7 0 0.00
8 0 0.01
9 0 0.06
10 0 0.04
11 0 0.14
12 0 0.01
13 0 0.01
14 0 0.00
15 0 0.06
16 0 0.00
0 1 0.03
1 1 0.62
2 1 0.01
3 1 0.01
4 1 0.00
5 1 0.00
6 1 0.01
7 1 0.08
8 1 0.03
9 1 0.01
10 1 0.01
11 1 0.09
12 1 0.03
13 1 0.03
14 1 0.01
15 1 0.01
16 1 0.01
0 2 0.00
1 2 0.01
2 2 0.77
3 2 0.02
4 2 0.01
5 2 0.01
6 2 0.03
7 2 0.00
8 2 0.03
9 2 0.00
10 2 0.01
11 2 0.01
12 2 0.00
13 2 0.06
14 2 0.03
15 2 0.00
16 2 0.04
0 3 0.03
1 3 0.01
2 3 0.01
3 3 0.66
4 3 0.01
5 3 0.00
6 3 0.01
7 3 0.00
8 3 0.01
9 3 0.04
10 3 0.01
11 3 0.11
12 3 0.01
13 3 0.04
14 3 0.01
15 3 0.04
16 3 0.01
0 4 0.01
1 4 0.00
2 4 0.01
3 4 0.03
4 4 0.73
5 4 0.01
6 4 0.04
7 4 0.00
8 4 0.00
9 4 0.01
10 4 0.01
11 4 0.10
12 4 0.01
13 4 0.03
14 4 0.01
15 4 0.01
16 4 0.01
0 5 0.00
1 5 0.00
2 5 0.01
3 5 0.01
4 5 0.01
5 5 0.95
6 5 0.01
7 5 0.00
8 5 0.00
9 5 0.01
10 5 0.00
11 5 0.01
12 5 0.00
13 5 0.01
14 5 0.00
15 5 0.00
16 5 0.00
0 6 0.01
1 6 0.01
2 6 0.22
3 6 0.04
4 6 0.01
5 6 0.00
6 6 0.36
7 6 0.00
8 6 0.01
9 6 0.03
10 6 0.01
11 6 0.12
12 6 0.00
13 6 0.10
14 6 0.04
15 6 0.01
16 6 0.03
0 7 0.00
1 7 0.01
2 7 0.00
3 7 0.00
4 7 0.00
5 7 0.00
6 7 0.00
7 7 0.96
8 7 0.00
9 7 0.00
10 7 0.00
11 7 0.00
12 7 0.01
13 7 0.00
14 7 0.02
15 7 0.00
16 7 0.00
0 8 0.00
1 8 0.01
2 8 0.00
3 8 0.01
4 8 0.00
5 8 0.00
6 8 0.01
7 8 0.00
8 8 0.74
9 8 0.01
10 8 0.00
11 8 0.00
12 8 0.00
13 8 0.07
14 8 0.01
15 8 0.00
16 8 0.16
0 9 0.07
1 9 0.02
2 9 0.01
3 9 0.04
4 9 0.01
5 9 0.00
6 9 0.01
7 9 0.01
8 9 0.01
9 9 0.54
10 9 0.04
11 9 0.14
12 9 0.02
13 9 0.02
14 9 0.01
15 9 0.05
16 9 0.00
0 10 0.01
1 10 0.01
2 10 0.01
3 10 0.01
4 10 0.01
5 10 0.00
6 10 0.04
7 10 0.01
8 10 0.00
9 10 0.01
10 10 0.66
11 10 0.11
12 10 0.03
13 10 0.04
14 10 0.01
15 10 0.02
16 10 0.01
0 11 0.04
1 11 0.03
2 11 0.01
3 11 0.04
4 11 0.04
5 11 0.00
6 11 0.03
7 11 0.01
8 11 0.01
9 11 0.04
10 11 0.03
11 11 0.68
12 11 0.03
13 11 0.01
14 11 0.00
15 11 0.01
16 11 0.00
0 12 0.01
1 12 0.01
2 12 0.00
3 12 0.03
4 12 0.03
5 12 0.00
6 12 0.01
7 12 0.00
8 12 0.00
9 12 0.03
10 12 0.01
11 12 0.06
12 12 0.79
13 12 0.01
14 12 0.01
15 12 0.01
16 12 0.00
0 13 0.01
1 13 0.02
2 13 0.04
3 13 0.06
4 13 0.01
5 13 0.01
6 13 0.03
7 13 0.00
8 13 0.05
9 13 0.01
10 13 0.00
11 13 0.04
12 13 0.01
13 13 0.64
14 13 0.04
15 13 0.00
16 13 0.06
0 14 0.00
1 14 0.01
2 14 0.00
3 14 0.01
4 14 0.01
5 14 0.00
6 14 0.01
7 14 0.01
8 14 0.00
9 14 0.00
10 14 0.01
11 14 0.01
12 14 0.00
13 14 0.06
14 14 0.89
15 14 0.00
16 14 0.00
0 15 0.07
1 15 0.03
2 15 0.01
3 15 0.02
4 15 0.01
5 15 0.00
6 15 0.01
7 15 0.00
8 15 0.00
9 15 0.04
10 15 0.02
11 15 0.16
12 15 0.01
13 15 0.01
14 15 0.00
15 15 0.62
16 15 0.00
0 16 0.00
1 16 0.01
2 16 0.01
3 16 0.01
4 16 0.00
5 16 0.00
6 16 0.01
7 16 0.00
8 16 0.17
9 16 0.00
10 16 0.00
11 16 0.01
12 16 0.00
13 16 0.06
14 16 0.04
15 16 0.00
16 16 0.69
};
\end{axis}
%\draw[black,thick] (2.625,5.8) circle(0.5);
%\draw[red,dashed,thick] (3.675,5.8) circle(0.5);
\end{tikzpicture}
\caption{Entropy KNN HOG confusion matrix (accuracy~0.6921)}\label{fig:conf_KHE}
\end{figure}

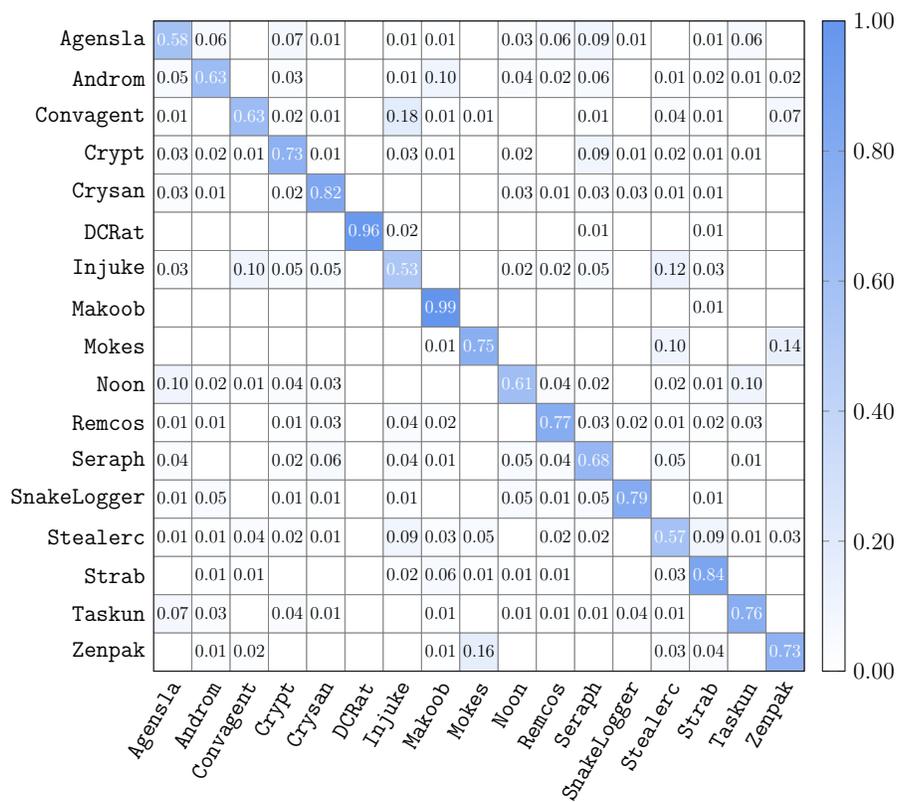
\begin{figure}[!htb]
\centering
%\begin{tikzpicture}[scale=0.8,every node/.style={scale=0.8}]
\begin{tikzpicture}[scale=0.60]
    \begin{axis}[%colorbar/width=2.5mm,
        width=16cm,
        height=16cm,
%        xlabel={\LARGE Predicted class},
%        ylabel={\LARGE Actual class},
%        colormap={blackwhite}{gray(0cm)=(1); gray(1cm)=(0.5)},
%	colormap={bluewhite}{color=(white) color=(blue)},
%	colormap={bluewhite}{color=(white) rgb255=(0,191,255)},
	colormap={bluewhite}{color=(white) rgb255=(100,149,237)},
        xticklabels={
        Agensla,
        Androm,
        Convagent,
        Crypt,
        Crysan,
        DCRat,
        Injuke,
        Makoob,
        Mokes,
        Noon,
        Remcos,
        Seraph,
        SnakeLogger,
        Stealerc,
        Strab,
        Taskun,
        Zenpak
        },
        xtick={0,...,16},
        xtick style={draw=none},
	xticklabel style={scale=1.25,anchor=east,rotate=60,yshift=-5pt,font=\tt},
        yticklabels={
        Agensla,
        Androm,
        Convagent,
        Crypt,
        Crysan,
        DCRat,
        Injuke,
        Makoob,
        Mokes,
        Noon,
        Remcos,
        Seraph,
        SnakeLogger,
        Stealerc,
        Strab,
        Taskun,
        Zenpak
        },
        ytick={0,...,16},
        ytick style={draw=none},
        enlargelimits=false,
        yticklabel style={scale=1.25,font=\tt},
        colorbar,
        colorbar style={
%     	  	width=0.05*\pgfkeysvalueof{/pgfplots/parent axis width},%%% added this
%     	  	height=0.5*\pgfkeysvalueof{/pgfplots/parent axis height},
%		plot graphics/node/.style={scale=1.33,anchor=south west,inner sep=0pt,}, %%% scale colorbar fill %%%
            ytick={0.00,0.20,0.40,0.60,0.80,1.00},
            yticklabels={0.00,0.20,0.40,0.60,0.80,1.00},
            yticklabel={\pgfmathprintnumber\tick},
            yticklabel style={%font=\footnotesize,
            		scale=1.25,
            		/pgf/number format/fixed,
			/pgf/number format/fixed zerofill,
			/pgf/number format/precision=2}
        },
%        point meta min=0,
%        point meta max=200,
        point meta min=0.0,
        point meta max=1.0,
        nodes near coords={\pgfmathprintnumber\pgfplotspointmeta},
        % ---------------------------------------------------------------------
        % show `nodes near coords' but adapt the style so that values
        % above a threshold get another style
        % (adapted from <http://tex.stackexchange.com/a/141006/95441>)
        % #1: the THRESHOLD after which we switch to a special display.
        nodes near coords black white/.style={
            % define the style of the nodes with "small" values
            small value/.style={
                yshift=-7pt,
%                text=white,
                text=black,
                /pgf/number format/fixed,
%                /pgf/number format/precision=0,
                /pgf/number format/precision=2,
                /pgf/number format/zerofill=true,
                scale=0.95,
%                /pgf/number format/precision=0
            },
            % define the style of the nodes with "large" values
            large value/.style={
                yshift=-7pt,
%                text=black,
                text=white,
                /pgf/number format/fixed,
%                /pgf/number format/precision=2,
                /pgf/number format/precision=2,
                /pgf/number format/zerofill=true,
                scale=0.95,
%                /pgf/number format/precision=0
            },
            every node near coord/.style={
                check for zero/.code={
                    \pgfmathfloatifflags{\pgfplotspointmeta}{0}{
                        % If meta=0, make the node a coordinate
                        % (which doesn't have text)
                        \pgfkeys{/tikz/coordinate}
                    }{
                        \begingroup
                        % this group is merely to switch to FPU locally.
                        % Might be unnecessary, but who knows.
                        \pgfkeys{/pgf/fpu}
                        \pgfmathparse{\pgfplotspointmeta<#1}
                        \global\let\result=\pgfmathresult
                        \endgroup
                        %
                        % simplifies debugging:
                        %\show\result
                        %
                        \pgfmathfloatcreate{1}{1.0}{0}
                        \let\ONE=\pgfmathresult
                        \ifx\result\ONE
                            % AH: our condition 'y < #1' is met.
                            \pgfkeysalso{/pgfplots/small value}
                        \else
                            % ok, proceed as usual.
                            \pgfkeysalso{/pgfplots/large value}
                        \fi
                    }
                },
                check for zero,
            },
        },
        % asign a value to the new style which is the threshold at which
        % the two style `small value' or `large value' are used
%        nodes near coords black white=100,
        nodes near coords black white=0.5,
        % -----------------------------------------------------------------
    ]
\addplot[
  matrix plot,
  mesh/cols=17,
  point meta=explicit,
  draw=gray
] table [x=x, y=y, meta=C] {
x y C
0 0 0.58
1 0 0.06
2 0 0.00
3 0 0.07
4 0 0.01
5 0 0.00
6 0 0.01
7 0 0.01
8 0 0.00
9 0 0.03
10 0 0.06
11 0 0.09
12 0 0.01
13 0 0.00
14 0 0.01
15 0 0.06
16 0 0.00
0 1 0.05
1 1 0.63
2 1 0.00
3 1 0.03
4 1 0.00
5 1 0.00
6 1 0.01
7 1 0.10
8 1 0.00
9 1 0.04
10 1 0.02
11 1 0.06
12 1 0.00
13 1 0.01
14 1 0.02
15 1 0.01
16 1 0.02
0 2 0.01
1 2 0.00
2 2 0.63
3 2 0.02
4 2 0.01
5 2 0.00
6 2 0.18
7 2 0.01
8 2 0.01
9 2 0.00
10 2 0.00
11 2 0.01
12 2 0.00
13 2 0.04
14 2 0.01
15 2 0.00
16 2 0.07
0 3 0.03
1 3 0.02
2 3 0.01
3 3 0.73
4 3 0.01
5 3 0.00
6 3 0.03
7 3 0.01
8 3 0.00
9 3 0.02
10 3 0.00
11 3 0.09
12 3 0.01
13 3 0.02
14 3 0.01
15 3 0.01
16 3 0.00
0 4 0.03
1 4 0.01
2 4 0.00
3 4 0.02
4 4 0.82
5 4 0.00
6 4 0.00
7 4 0.00
8 4 0.00
9 4 0.03
10 4 0.01
11 4 0.03
12 4 0.03
13 4 0.01
14 4 0.01
15 4 0.00
16 4 0.00
0 5 0.00
1 5 0.00
2 5 0.00
3 5 0.00
4 5 0.00
5 5 0.96
6 5 0.02
7 5 0.00
8 5 0.00
9 5 0.00
10 5 0.00
11 5 0.01
12 5 0.00
13 5 0.00
14 5 0.01
15 5 0.00
16 5 0.00
0 6 0.03
1 6 0.00
2 6 0.10
3 6 0.05
4 6 0.05
5 6 0.00
6 6 0.53
7 6 0.00
8 6 0.00
9 6 0.02
10 6 0.02
11 6 0.05
12 6 0.00
13 6 0.12
14 6 0.03
15 6 0.00
16 6 0.00
0 7 0.00
1 7 0.00
2 7 0.00
3 7 0.00
4 7 0.00
5 7 0.00
6 7 0.00
7 7 0.99
8 7 0.00
9 7 0.00
10 7 0.00
11 7 0.00
12 7 0.00
13 7 0.00
14 7 0.01
15 7 0.00
16 7 0.00
0 8 0.00
1 8 0.00
2 8 0.00
3 8 0.00
4 8 0.00
5 8 0.00
6 8 0.00
7 8 0.01
8 8 0.75
9 8 0.00
10 8 0.00
11 8 0.00
12 8 0.00
13 8 0.10
14 8 0.00
15 8 0.00
16 8 0.14
0 9 0.10
1 9 0.02
2 9 0.01
3 9 0.04
4 9 0.03
5 9 0.00
6 9 0.00
7 9 0.00
8 9 0.00
9 9 0.61
10 9 0.04
11 9 0.02
12 9 0.00
13 9 0.02
14 9 0.01
15 9 0.10
16 9 0.00
0 10 0.01
1 10 0.01
2 10 0.00
3 10 0.01
4 10 0.03
5 10 0.00
6 10 0.04
7 10 0.02
8 10 0.00
9 10 0.00
10 10 0.77
11 10 0.03
12 10 0.02
13 10 0.01
14 10 0.02
15 10 0.03
16 10 0.00
0 11 0.04
1 11 0.00
2 11 0.00
3 11 0.02
4 11 0.06
5 11 0.00
6 11 0.04
7 11 0.01
8 11 0.00
9 11 0.05
10 11 0.04
11 11 0.68
12 11 0.00
13 11 0.05
14 11 0.00
15 11 0.01
16 11 0.00
0 12 0.01
1 12 0.05
2 12 0.00
3 12 0.01
4 12 0.01
5 12 0.00
6 12 0.01
7 12 0.00
8 12 0.00
9 12 0.05
10 12 0.01
11 12 0.05
12 12 0.79
13 12 0.00
14 12 0.01
15 12 0.00
16 12 0.00
0 13 0.01
1 13 0.01
2 13 0.04
3 13 0.02
4 13 0.01
5 13 0.00
6 13 0.09
7 13 0.03
8 13 0.05
9 13 0.00
10 13 0.02
11 13 0.02
12 13 0.00
13 13 0.57
14 13 0.09
15 13 0.01
16 13 0.03
0 14 0.00
1 14 0.01
2 14 0.01
3 14 0.00
4 14 0.00
5 14 0.00
6 14 0.02
7 14 0.06
8 14 0.01
9 14 0.01
10 14 0.01
11 14 0.00
12 14 0.00
13 14 0.03
14 14 0.84
15 14 0.00
16 14 0.00
0 15 0.07
1 15 0.03
2 15 0.00
3 15 0.04
4 15 0.01
5 15 0.00
6 15 0.00
7 15 0.01
8 15 0.00
9 15 0.01
10 15 0.01
11 15 0.01
12 15 0.04
13 15 0.01
14 15 0.00
15 15 0.76
16 15 0.00
0 16 0.00
1 16 0.01
2 16 0.02
3 16 0.00
4 16 0.00
5 16 0.00
6 16 0.00
7 16 0.01
8 16 0.16
9 16 0.00
10 16 0.00
11 16 0.00
12 16 0.00
13 16 0.03
14 16 0.04
15 16 0.00
16 16 0.73
};
\end{axis}
%\draw[black,thick] (2.625,5.8) circle(0.5);
%\draw[red,dashed,thick] (3.675,5.8) circle(0.5);
\end{tikzpicture}
\caption{Byteclass XGBoost Haralick confusion matrix (accuracy~0.7147)}\label{fig:conf_XHB}
\end{figure}

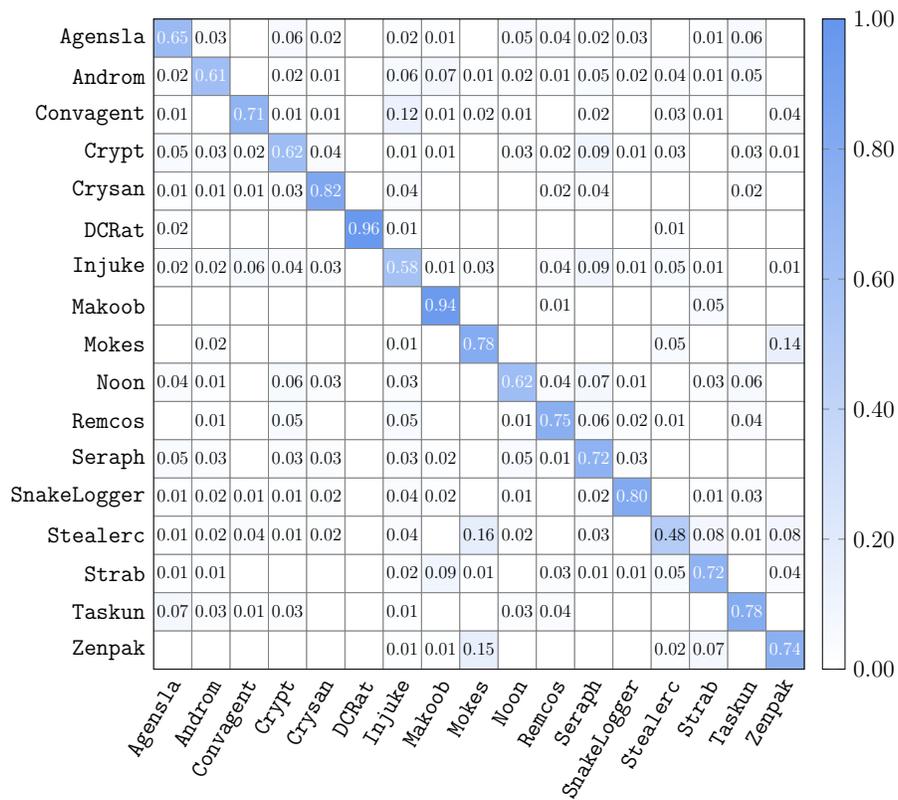
\begin{figure}[!htb]
\centering
%\begin{tikzpicture}[scale=0.8,every node/.style={scale=0.8}]
\begin{tikzpicture}[scale=0.60]
    \begin{axis}[%colorbar/width=2.5mm,
        width=16cm,
        height=16cm,
%        xlabel={\LARGE Predicted class},
%        ylabel={\LARGE Actual class},
%        colormap={blackwhite}{gray(0cm)=(1); gray(1cm)=(0.5)},
%	colormap={bluewhite}{color=(white) color=(blue)},
%	colormap={bluewhite}{color=(white) rgb255=(0,191,255)},
	colormap={bluewhite}{color=(white) rgb255=(100,149,237)},
        xticklabels={
        Agensla,
        Androm,
        Convagent,
        Crypt,
        Crysan,
        DCRat,
        Injuke,
        Makoob,
        Mokes,
        Noon,
        Remcos,
        Seraph,
        SnakeLogger,
        Stealerc,
        Strab,
        Taskun,
        Zenpak
        },
        xtick={0,...,16},
        xtick style={draw=none},
	xticklabel style={scale=1.25,anchor=east,rotate=60,yshift=-5pt,font=\tt},
        yticklabels={
        Agensla,
        Androm,
        Convagent,
        Crypt,
        Crysan,
        DCRat,
        Injuke,
        Makoob,
        Mokes,
        Noon,
        Remcos,
        Seraph,
        SnakeLogger,
        Stealerc,
        Strab,
        Taskun,
        Zenpak
        },
        ytick={0,...,16},
        ytick style={draw=none},
        enlargelimits=false,
        yticklabel style={scale=1.25,font=\tt},
        colorbar,
        colorbar style={
%     	  	width=0.05*\pgfkeysvalueof{/pgfplots/parent axis width},%%% added this
%     	  	height=0.5*\pgfkeysvalueof{/pgfplots/parent axis height},
%		plot graphics/node/.style={scale=1.33,anchor=south west,inner sep=0pt,}, %%% scale colorbar fill %%%
            ytick={0.00,0.20,0.40,0.60,0.80,1.00},
            yticklabels={0.00,0.20,0.40,0.60,0.80,1.00},
            yticklabel={\pgfmathprintnumber\tick},
            yticklabel style={%font=\footnotesize,
            		scale=1.25,
            		/pgf/number format/fixed,
			/pgf/number format/fixed zerofill,
			/pgf/number format/precision=2}
        },
%        point meta min=0,
%        point meta max=200,
        point meta min=0.0,
        point meta max=1.0,
        nodes near coords={\pgfmathprintnumber\pgfplotspointmeta},
        % ---------------------------------------------------------------------
        % show `nodes near coords' but adapt the style so that values
        % above a threshold get another style
        % (adapted from <http://tex.stackexchange.com/a/141006/95441>)
        % #1: the THRESHOLD after which we switch to a special display.
        nodes near coords black white/.style={
            % define the style of the nodes with "small" values
            small value/.style={
                yshift=-7pt,
%                text=white,
                text=black,
                /pgf/number format/fixed,
%                /pgf/number format/precision=0,
                /pgf/number format/precision=2,
                /pgf/number format/zerofill=true,
                scale=0.95,
%                /pgf/number format/precision=0
            },
            % define the style of the nodes with "large" values
            large value/.style={
                yshift=-7pt,
%                text=black,
                text=white,
                /pgf/number format/fixed,
%                /pgf/number format/precision=2,
                /pgf/number format/precision=2,
                /pgf/number format/zerofill=true,
                scale=0.95,
%                /pgf/number format/precision=0
            },
            every node near coord/.style={
                check for zero/.code={
                    \pgfmathfloatifflags{\pgfplotspointmeta}{0}{
                        % If meta=0, make the node a coordinate
                        % (which doesn't have text)
                        \pgfkeys{/tikz/coordinate}
                    }{
                        \begingroup
                        % this group is merely to switch to FPU locally.
                        % Might be unnecessary, but who knows.
                        \pgfkeys{/pgf/fpu}
                        \pgfmathparse{\pgfplotspointmeta<#1}
                        \global\let\result=\pgfmathresult
                        \endgroup
                        %
                        % simplifies debugging:
                        %\show\result
                        %
                        \pgfmathfloatcreate{1}{1.0}{0}
                        \let\ONE=\pgfmathresult
                        \ifx\result\ONE
                            % AH: our condition 'y < #1' is met.
                            \pgfkeysalso{/pgfplots/small value}
                        \else
                            % ok, proceed as usual.
                            \pgfkeysalso{/pgfplots/large value}
                        \fi
                    }
                },
                check for zero,
            },
        },
        % asign a value to the new style which is the threshold at which
        % the two style `small value' or `large value' are used
%        nodes near coords black white=100,
        nodes near coords black white=0.5,
        % -----------------------------------------------------------------
    ]
\addplot[
  matrix plot,
  mesh/cols=17,
  point meta=explicit,
  draw=gray
] table [x=x, y=y, meta=C] {
x y C
0 0 0.65
1 0 0.03
2 0 0.00
3 0 0.06
4 0 0.02
5 0 0.00
6 0 0.02
7 0 0.01
8 0 0.00
9 0 0.05
10 0 0.04
11 0 0.02
12 0 0.03
13 0 0.00
14 0 0.01
15 0 0.06
16 0 0.00
0 1 0.02
1 1 0.61
2 1 0.00
3 1 0.02
4 1 0.01
5 1 0.00
6 1 0.06
7 1 0.07
8 1 0.01
9 1 0.02
10 1 0.01
11 1 0.05
12 1 0.02
13 1 0.04
14 1 0.01
15 1 0.05
16 1 0.00
0 2 0.01
1 2 0.00
2 2 0.71
3 2 0.01
4 2 0.01
5 2 0.00
6 2 0.12
7 2 0.01
8 2 0.02
9 2 0.01
10 2 0.00
11 2 0.02
12 2 0.00
13 2 0.03
14 2 0.01
15 2 0.00
16 2 0.04
0 3 0.05
1 3 0.03
2 3 0.02
3 3 0.62
4 3 0.04
5 3 0.00
6 3 0.01
7 3 0.01
8 3 0.00
9 3 0.03
10 3 0.02
11 3 0.09
12 3 0.01
13 3 0.03
14 3 0.00
15 3 0.03
16 3 0.01
0 4 0.01
1 4 0.01
2 4 0.01
3 4 0.03
4 4 0.82
5 4 0.00
6 4 0.04
7 4 0.00
8 4 0.00
9 4 0.00
10 4 0.02
11 4 0.04
12 4 0.00
13 4 0.00
14 4 0.00
15 4 0.02
16 4 0.00
0 5 0.02
1 5 0.00
2 5 0.00
3 5 0.00
4 5 0.00
5 5 0.96
6 5 0.01
7 5 0.00
8 5 0.00
9 5 0.00
10 5 0.00
11 5 0.00
12 5 0.00
13 5 0.01
14 5 0.00
15 5 0.00
16 5 0.00
0 6 0.02
1 6 0.02
2 6 0.06
3 6 0.04
4 6 0.03
5 6 0.00
6 6 0.58
7 6 0.01
8 6 0.03
9 6 0.00
10 6 0.04
11 6 0.09
12 6 0.01
13 6 0.05
14 6 0.01
15 6 0.00
16 6 0.01
0 7 0.00
1 7 0.00
2 7 0.00
3 7 0.00
4 7 0.00
5 7 0.00
6 7 0.00
7 7 0.94
8 7 0.00
9 7 0.00
10 7 0.01
11 7 0.00
12 7 0.00
13 7 0.00
14 7 0.05
15 7 0.00
16 7 0.00
0 8 0.00
1 8 0.02
2 8 0.00
3 8 0.00
4 8 0.00
5 8 0.00
6 8 0.01
7 8 0.00
8 8 0.78
9 8 0.00
10 8 0.00
11 8 0.00
12 8 0.00
13 8 0.05
14 8 0.00
15 8 0.00
16 8 0.14
0 9 0.04
1 9 0.01
2 9 0.00
3 9 0.06
4 9 0.03
5 9 0.00
6 9 0.03
7 9 0.00
8 9 0.00
9 9 0.62
10 9 0.04
11 9 0.07
12 9 0.01
13 9 0.00
14 9 0.03
15 9 0.06
16 9 0.00
0 10 0.00
1 10 0.01
2 10 0.00
3 10 0.05
4 10 0.00
5 10 0.00
6 10 0.05
7 10 0.00
8 10 0.00
9 10 0.01
10 10 0.75
11 10 0.06
12 10 0.02
13 10 0.01
14 10 0.00
15 10 0.04
16 10 0.00
0 11 0.05
1 11 0.03
2 11 0.00
3 11 0.03
4 11 0.03
5 11 0.00
6 11 0.03
7 11 0.02
8 11 0.00
9 11 0.05
10 11 0.01
11 11 0.72
12 11 0.03
13 11 0.00
14 11 0.00
15 11 0.00
16 11 0.00
0 12 0.01
1 12 0.02
2 12 0.01
3 12 0.01
4 12 0.02
5 12 0.00
6 12 0.04
7 12 0.02
8 12 0.00
9 12 0.01
10 12 0.00
11 12 0.02
12 12 0.80
13 12 0.00
14 12 0.01
15 12 0.03
16 12 0.00
0 13 0.01
1 13 0.02
2 13 0.04
3 13 0.01
4 13 0.02
5 13 0.00
6 13 0.04
7 13 0.00
8 13 0.16
9 13 0.02
10 13 0.00
11 13 0.03
12 13 0.00
13 13 0.48
14 13 0.08
15 13 0.01
16 13 0.08
0 14 0.01
1 14 0.01
2 14 0.00
3 14 0.00
4 14 0.00
5 14 0.00
6 14 0.02
7 14 0.09
8 14 0.01
9 14 0.00
10 14 0.03
11 14 0.01
12 14 0.01
13 14 0.05
14 14 0.72
15 14 0.00
16 14 0.04
0 15 0.07
1 15 0.03
2 15 0.01
3 15 0.03
4 15 0.00
5 15 0.00
6 15 0.01
7 15 0.00
8 15 0.00
9 15 0.03
10 15 0.04
11 15 0.00
12 15 0.00
13 15 0.00
14 15 0.00
15 15 0.78
16 15 0.00
0 16 0.00
1 16 0.00
2 16 0.00
3 16 0.00
4 16 0.00
5 16 0.00
6 16 0.01
7 16 0.01
8 16 0.15
9 16 0.00
10 16 0.00
11 16 0.00
12 16 0.00
13 16 0.02
14 16 0.07
15 16 0.00
16 16 0.74
};
\end{axis}
%\draw[black,thick] (2.625,5.8) circle(0.5);
%\draw[red,dashed,thick] (3.675,5.8) circle(0.5);
\end{tikzpicture}
\caption{Hilbert XGBoost Haralick confusion matrix (accuracy~0.7100)}\label{fig:conf_XHBH}
\end{figure}

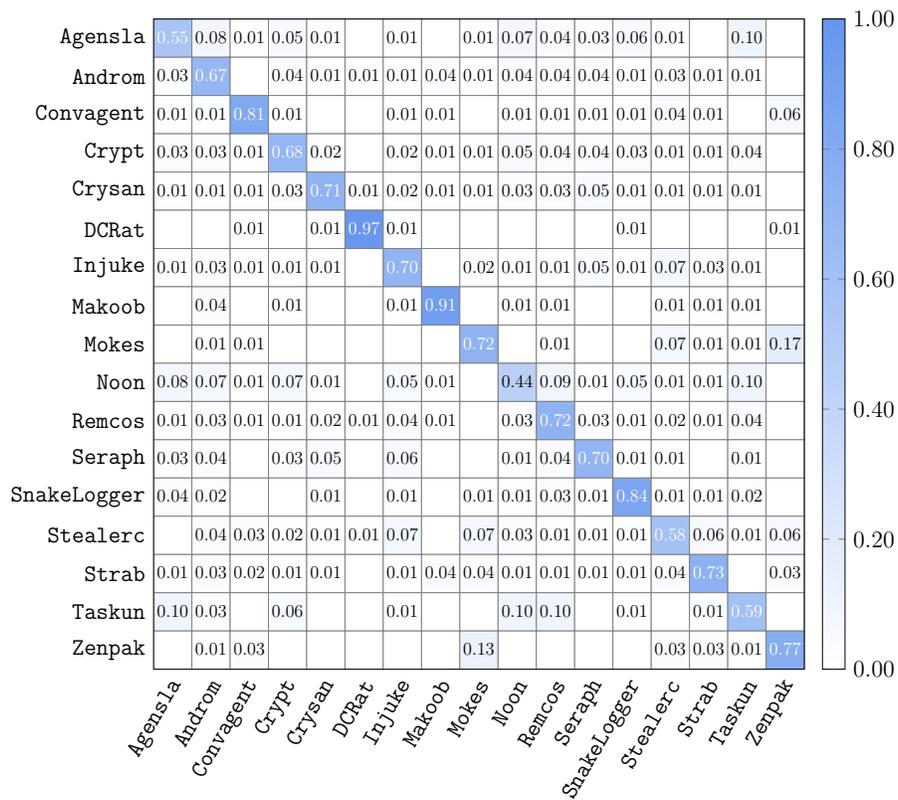
\begin{figure}[!htb]
\centering
%\begin{tikzpicture}[scale=0.8,every node/.style={scale=0.8}]
\begin{tikzpicture}[scale=0.60]
    \begin{axis}[%colorbar/width=2.5mm,
        width=16cm,
        height=16cm,
%        xlabel={\LARGE Predicted class},
%        ylabel={\LARGE Actual class},
%        colormap={blackwhite}{gray(0cm)=(1); gray(1cm)=(0.5)},
%	colormap={bluewhite}{color=(white) color=(blue)},
%	colormap={bluewhite}{color=(white) rgb255=(0,191,255)},
	colormap={bluewhite}{color=(white) rgb255=(100,149,237)},
        xticklabels={
        Agensla,
        Androm,
        Convagent,
        Crypt,
        Crysan,
        DCRat,
        Injuke,
        Makoob,
        Mokes,
        Noon,
        Remcos,
        Seraph,
        SnakeLogger,
        Stealerc,
        Strab,
        Taskun,
        Zenpak
        },
        xtick={0,...,16},
        xtick style={draw=none},
	xticklabel style={scale=1.25,anchor=east,rotate=60,yshift=-5pt,font=\tt},
        yticklabels={
        Agensla,
        Androm,
        Convagent,
        Crypt,
        Crysan,
        DCRat,
        Injuke,
        Makoob,
        Mokes,
        Noon,
        Remcos,
        Seraph,
        SnakeLogger,
        Stealerc,
        Strab,
        Taskun,
        Zenpak
        },
        ytick={0,...,16},
        ytick style={draw=none},
        enlargelimits=false,
        yticklabel style={scale=1.25,font=\tt},
        colorbar,
        colorbar style={
%     	  	width=0.05*\pgfkeysvalueof{/pgfplots/parent axis width},%%% added this
%     	  	height=0.5*\pgfkeysvalueof{/pgfplots/parent axis height},
%		plot graphics/node/.style={scale=1.33,anchor=south west,inner sep=0pt,}, %%% scale colorbar fill %%%
            ytick={0.00,0.20,0.40,0.60,0.80,1.00},
            yticklabels={0.00,0.20,0.40,0.60,0.80,1.00},
            yticklabel={\pgfmathprintnumber\tick},
            yticklabel style={%font=\footnotesize,
            		scale=1.25,
            		/pgf/number format/fixed,
			/pgf/number format/fixed zerofill,
			/pgf/number format/precision=2}
        },
%        point meta min=0,
%        point meta max=200,
        point meta min=0.0,
        point meta max=1.0,
        nodes near coords={\pgfmathprintnumber\pgfplotspointmeta},
        % ---------------------------------------------------------------------
        % show `nodes near coords' but adapt the style so that values
        % above a threshold get another style
        % (adapted from <http://tex.stackexchange.com/a/141006/95441>)
        % #1: the THRESHOLD after which we switch to a special display.
        nodes near coords black white/.style={
            % define the style of the nodes with "small" values
            small value/.style={
                yshift=-7pt,
%                text=white,
                text=black,
                /pgf/number format/fixed,
%                /pgf/number format/precision=0,
                /pgf/number format/precision=2,
                /pgf/number format/zerofill=true,
                scale=0.95,
%                /pgf/number format/precision=0
            },
            % define the style of the nodes with "large" values
            large value/.style={
                yshift=-7pt,
%                text=black,
                text=white,
                /pgf/number format/fixed,
%                /pgf/number format/precision=2,
                /pgf/number format/precision=2,
                /pgf/number format/zerofill=true,
                scale=0.95,
%                /pgf/number format/precision=0
            },
            every node near coord/.style={
                check for zero/.code={
                    \pgfmathfloatifflags{\pgfplotspointmeta}{0}{
                        % If meta=0, make the node a coordinate
                        % (which doesn't have text)
                        \pgfkeys{/tikz/coordinate}
                    }{
                        \begingroup
                        % this group is merely to switch to FPU locally.
                        % Might be unnecessary, but who knows.
                        \pgfkeys{/pgf/fpu}
                        \pgfmathparse{\pgfplotspointmeta<#1}
                        \global\let\result=\pgfmathresult
                        \endgroup
                        %
                        % simplifies debugging:
                        %\show\result
                        %
                        \pgfmathfloatcreate{1}{1.0}{0}
                        \let\ONE=\pgfmathresult
                        \ifx\result\ONE
                            % AH: our condition 'y < #1' is met.
                            \pgfkeysalso{/pgfplots/small value}
                        \else
                            % ok, proceed as usual.
                            \pgfkeysalso{/pgfplots/large value}
                        \fi
                    }
                },
                check for zero,
            },
        },
        % asign a value to the new style which is the threshold at which
        % the two style `small value' or `large value' are used
%        nodes near coords black white=100,
        nodes near coords black white=0.5,
        % -----------------------------------------------------------------
    ]
\addplot[
  matrix plot,
  mesh/cols=17,
  point meta=explicit,
  draw=gray
] table [x=x, y=y, meta=C] {
x y C
0 0 0.55
1 0 0.08
2 0 0.01
3 0 0.05
4 0 0.01
5 0 0.00
6 0 0.01
7 0 0.00
8 0 0.01
9 0 0.07
10 0 0.04
11 0 0.03
12 0 0.06
13 0 0.01
14 0 0.00
15 0 0.10
16 0 0.00
0 1 0.03
1 1 0.67
2 1 0.00
3 1 0.04
4 1 0.01
5 1 0.01
6 1 0.01
7 1 0.04
8 1 0.01
9 1 0.04
10 1 0.04
11 1 0.04
12 1 0.01
13 1 0.03
14 1 0.01
15 1 0.01
16 1 0.00
0 2 0.01
1 2 0.01
2 2 0.81
3 2 0.01
4 2 0.00
5 2 0.00
6 2 0.01
7 2 0.01
8 2 0.00
9 2 0.01
10 2 0.01
11 2 0.01
12 2 0.01
13 2 0.04
14 2 0.01
15 2 0.00
16 2 0.06
0 3 0.03
1 3 0.03
2 3 0.01
3 3 0.68
4 3 0.02
5 3 0.00
6 3 0.02
7 3 0.01
8 3 0.01
9 3 0.05
10 3 0.04
11 3 0.04
12 3 0.03
13 3 0.01
14 3 0.01
15 3 0.04
16 3 0.00
0 4 0.01
1 4 0.01
2 4 0.01
3 4 0.03
4 4 0.71
5 4 0.01
6 4 0.02
7 4 0.01
8 4 0.01
9 4 0.03
10 4 0.03
11 4 0.05
12 4 0.01
13 4 0.01
14 4 0.01
15 4 0.01
16 4 0.00
0 5 0.00
1 5 0.00
2 5 0.01
3 5 0.00
4 5 0.01
5 5 0.97
6 5 0.01
7 5 0.00
8 5 0.00
9 5 0.00
10 5 0.00
11 5 0.00
12 5 0.01
13 5 0.00
14 5 0.00
15 5 0.00
16 5 0.01
0 6 0.01
1 6 0.03
2 6 0.01
3 6 0.01
4 6 0.01
5 6 0.00
6 6 0.70
7 6 0.00
8 6 0.02
9 6 0.01
10 6 0.01
11 6 0.05
12 6 0.01
13 6 0.07
14 6 0.03
15 6 0.01
16 6 0.00
0 7 0.00
1 7 0.04
2 7 0.00
3 7 0.01
4 7 0.00
5 7 0.00
6 7 0.01
7 7 0.91
8 7 0.00
9 7 0.01
10 7 0.01
11 7 0.00
12 7 0.00
13 7 0.01
14 7 0.01
15 7 0.01
16 7 0.00
0 8 0.00
1 8 0.01
2 8 0.01
3 8 0.00
4 8 0.00
5 8 0.00
6 8 0.00
7 8 0.00
8 8 0.72
9 8 0.00
10 8 0.01
11 8 0.00
12 8 0.00
13 8 0.07
14 8 0.01
15 8 0.01
16 8 0.17
0 9 0.08
1 9 0.07
2 9 0.01
3 9 0.07
4 9 0.01
5 9 0.00
6 9 0.05
7 9 0.01
8 9 0.00
9 9 0.44
10 9 0.09
11 9 0.01
12 9 0.05
13 9 0.01
14 9 0.01
15 9 0.10
16 9 0.00
0 10 0.01
1 10 0.03
2 10 0.01
3 10 0.01
4 10 0.02
5 10 0.01
6 10 0.04
7 10 0.01
8 10 0.00
9 10 0.03
10 10 0.72
11 10 0.03
12 10 0.01
13 10 0.02
14 10 0.01
15 10 0.04
16 10 0.00
0 11 0.03
1 11 0.04
2 11 0.00
3 11 0.03
4 11 0.05
5 11 0.00
6 11 0.06
7 11 0.00
8 11 0.00
9 11 0.01
10 11 0.04
11 11 0.70
12 11 0.01
13 11 0.01
14 11 0.00
15 11 0.01
16 11 0.00
0 12 0.04
1 12 0.02
2 12 0.00
3 12 0.00
4 12 0.01
5 12 0.00
6 12 0.01
7 12 0.00
8 12 0.01
9 12 0.01
10 12 0.03
11 12 0.01
12 12 0.84
13 12 0.01
14 12 0.01
15 12 0.02
16 12 0.00
0 13 0.00
1 13 0.04
2 13 0.03
3 13 0.02
4 13 0.01
5 13 0.01
6 13 0.07
7 13 0.00
8 13 0.07
9 13 0.03
10 13 0.01
11 13 0.01
12 13 0.01
13 13 0.58
14 13 0.06
15 13 0.01
16 13 0.06
0 14 0.01
1 14 0.03
2 14 0.02
3 14 0.01
4 14 0.01
5 14 0.00
6 14 0.01
7 14 0.04
8 14 0.04
9 14 0.01
10 14 0.01
11 14 0.01
12 14 0.01
13 14 0.04
14 14 0.73
15 14 0.00
16 14 0.03
0 15 0.10
1 15 0.03
2 15 0.00
3 15 0.06
4 15 0.00
5 15 0.00
6 15 0.01
7 15 0.00
8 15 0.00
9 15 0.10
10 15 0.10
11 15 0.00
12 15 0.01
13 15 0.00
14 15 0.01
15 15 0.59
16 15 0.00
0 16 0.00
1 16 0.01
2 16 0.03
3 16 0.00
4 16 0.00
5 16 0.00
6 16 0.00
7 16 0.00
8 16 0.13
9 16 0.00
10 16 0.00
11 16 0.00
12 16 0.00
13 16 0.03
14 16 0.03
15 16 0.01
16 16 0.77
};
\end{axis}
%\draw[black,thick] (2.625,5.8) circle(0.5);
%\draw[red,dashed,thick] (3.675,5.8) circle(0.5);
\end{tikzpicture}
\caption{Cartesian KNN HOG confusion matrix (accuracy~0.7029)}\label{fig:conf_KHC}
\end{figure}

\begin{figure}[!htb]
\centering
%\begin{tikzpicture}[scale=0.8,every node/.style={scale=0.8}]
\begin{tikzpicture}[scale=0.60]
    \begin{axis}[%colorbar/width=2.5mm,
        width=16cm,
        height=16cm,
%        xlabel={\LARGE Predicted class},
%        ylabel={\LARGE Actual class},
%        colormap={blackwhite}{gray(0cm)=(1); gray(1cm)=(0.5)},
%	colormap={bluewhite}{color=(white) color=(blue)},
%	colormap={bluewhite}{color=(white) rgb255=(0,191,255)},
	colormap={bluewhite}{color=(white) rgb255=(100,149,237)},
        xticklabels={
        Agensla,
        Androm,
        Convagent,
        Crypt,
        Crysan,
        DCRat,
        Injuke,
        Makoob,
        Mokes,
        Noon,
        Remcos,
        Seraph,
        SnakeLogger,
        Stealerc,
        Strab,
        Taskun,
        Zenpak
        },
        xtick={0,...,16},
        xtick style={draw=none},
	xticklabel style={scale=1.25,anchor=east,rotate=60,yshift=-5pt,font=\tt},
        yticklabels={
        Agensla,
        Androm,
        Convagent,
        Crypt,
        Crysan,
        DCRat,
        Injuke,
        Makoob,
        Mokes,
        Noon,
        Remcos,
        Seraph,
        SnakeLogger,
        Stealerc,
        Strab,
        Taskun,
        Zenpak
        },
        ytick={0,...,16},
        ytick style={draw=none},
        enlargelimits=false,
        yticklabel style={scale=1.25,font=\tt},
        colorbar,
        colorbar style={
%     	  	width=0.05*\pgfkeysvalueof{/pgfplots/parent axis width},%%% added this
%     	  	height=0.5*\pgfkeysvalueof{/pgfplots/parent axis height},
%		plot graphics/node/.style={scale=1.33,anchor=south west,inner sep=0pt,}, %%% scale colorbar fill %%%
            ytick={0.00,0.20,0.40,0.60,0.80,1.00},
            yticklabels={0.00,0.20,0.40,0.60,0.80,1.00},
            yticklabel={\pgfmathprintnumber\tick},
            yticklabel style={%font=\footnotesize,
            		scale=1.25,
            		/pgf/number format/fixed,
			/pgf/number format/fixed zerofill,
			/pgf/number format/precision=2}
        },
%        point meta min=0,
%        point meta max=200,
        point meta min=0.0,
        point meta max=1.0,
        nodes near coords={\pgfmathprintnumber\pgfplotspointmeta},
        % ---------------------------------------------------------------------
        % show `nodes near coords' but adapt the style so that values
        % above a threshold get another style
        % (adapted from <http://tex.stackexchange.com/a/141006/95441>)
        % #1: the THRESHOLD after which we switch to a special display.
        nodes near coords black white/.style={
            % define the style of the nodes with "small" values
            small value/.style={
                yshift=-7pt,
%                text=white,
                text=black,
                /pgf/number format/fixed,
%                /pgf/number format/precision=0,
                /pgf/number format/precision=2,
                /pgf/number format/zerofill=true,
                scale=0.95,
%                /pgf/number format/precision=0
            },
            % define the style of the nodes with "large" values
            large value/.style={
                yshift=-7pt,
%                text=black,
                text=white,
                /pgf/number format/fixed,
%                /pgf/number format/precision=2,
                /pgf/number format/precision=2,
                /pgf/number format/zerofill=true,
                scale=0.95,
%                /pgf/number format/precision=0
            },
            every node near coord/.style={
                check for zero/.code={
                    \pgfmathfloatifflags{\pgfplotspointmeta}{0}{
                        % If meta=0, make the node a coordinate
                        % (which doesn't have text)
                        \pgfkeys{/tikz/coordinate}
                    }{
                        \begingroup
                        % this group is merely to switch to FPU locally.
                        % Might be unnecessary, but who knows.
                        \pgfkeys{/pgf/fpu}
                        \pgfmathparse{\pgfplotspointmeta<#1}
                        \global\let\result=\pgfmathresult
                        \endgroup
                        %
                        % simplifies debugging:
                        %\show\result
                        %
                        \pgfmathfloatcreate{1}{1.0}{0}
                        \let\ONE=\pgfmathresult
                        \ifx\result\ONE
                            % AH: our condition 'y < #1' is met.
                            \pgfkeysalso{/pgfplots/small value}
                        \else
                            % ok, proceed as usual.
                            \pgfkeysalso{/pgfplots/large value}
                        \fi
                    }
                },
                check for zero,
            },
        },
        % asign a value to the new style which is the threshold at which
        % the two style `small value' or `large value' are used
%        nodes near coords black white=100,
        nodes near coords black white=0.5,
        % -----------------------------------------------------------------
    ]
\addplot[
  matrix plot,
  mesh/cols=17,
  point meta=explicit,
  draw=gray
] table [x=x, y=y, meta=C] {
x y C
0 0 0.48
1 0 0.03
2 0 0.01
3 0 0.03
4 0 0.00
5 0 0.01
6 0 0.04
7 0 0.01
8 0 0.00
9 0 0.14
10 0 0.07
11 0 0.03
12 0 0.03
13 0 0.00
14 0 0.01
15 0 0.15
16 0 0.00
0 1 0.01
1 1 0.60
2 1 0.01
3 1 0.02
4 1 0.01
5 1 0.00
6 1 0.04
7 1 0.06
8 1 0.03
9 1 0.02
10 1 0.04
11 1 0.04
12 1 0.03
13 1 0.01
14 1 0.01
15 1 0.04
16 1 0.03
0 2 0.01
1 2 0.01
2 2 0.79
3 2 0.03
4 2 0.00
5 2 0.00
6 2 0.02
7 2 0.01
8 2 0.01
9 2 0.01
10 2 0.00
11 2 0.00
12 2 0.01
13 2 0.04
14 2 0.01
15 2 0.01
16 2 0.04
0 3 0.04
1 3 0.03
2 3 0.01
3 3 0.63
4 3 0.01
5 3 0.00
6 3 0.01
7 3 0.01
8 3 0.00
9 3 0.11
10 3 0.02
11 3 0.04
12 3 0.02
13 3 0.01
14 3 0.01
15 3 0.03
16 3 0.00
0 4 0.02
1 4 0.03
2 4 0.01
3 4 0.03
4 4 0.75
5 4 0.00
6 4 0.04
7 4 0.01
8 4 0.00
9 4 0.03
10 4 0.01
11 4 0.07
12 4 0.01
13 4 0.00
14 4 0.01
15 4 0.01
16 4 0.00
0 5 0.01
1 5 0.00
2 5 0.00
3 5 0.00
4 5 0.01
5 5 0.98
6 5 0.00
7 5 0.00
8 5 0.00
9 5 0.00
10 5 0.00
11 5 0.01
12 5 0.00
13 5 0.00
14 5 0.00
15 5 0.00
16 5 0.01
0 6 0.01
1 6 0.03
2 6 0.03
3 6 0.03
4 6 0.01
5 6 0.01
6 6 0.65
7 6 0.01
8 6 0.01
9 6 0.01
10 6 0.06
11 6 0.06
12 6 0.00
13 6 0.07
14 6 0.01
15 6 0.02
16 6 0.01
0 7 0.00
1 7 0.03
2 7 0.00
3 7 0.01
4 7 0.00
5 7 0.00
6 7 0.00
7 7 0.93
8 7 0.00
9 7 0.00
10 7 0.01
11 7 0.00
12 7 0.00
13 7 0.01
14 7 0.02
15 7 0.00
16 7 0.00
0 8 0.01
1 8 0.00
2 8 0.03
3 8 0.00
4 8 0.00
5 8 0.00
6 8 0.01
7 8 0.00
8 8 0.76
9 8 0.00
10 8 0.00
11 8 0.00
12 8 0.00
13 8 0.07
14 8 0.01
15 8 0.00
16 8 0.12
0 9 0.07
1 9 0.03
2 9 0.01
3 9 0.05
4 9 0.03
5 9 0.00
6 9 0.02
7 9 0.01
8 9 0.00
9 9 0.51
10 9 0.07
11 9 0.03
12 9 0.04
13 9 0.01
14 9 0.00
15 9 0.15
16 9 0.00
0 10 0.05
1 10 0.01
2 10 0.01
3 10 0.03
4 10 0.01
5 10 0.00
6 10 0.04
7 10 0.01
8 10 0.00
9 10 0.07
10 10 0.66
11 10 0.01
12 10 0.04
13 10 0.01
14 10 0.02
15 10 0.03
16 10 0.00
0 11 0.02
1 11 0.03
2 11 0.01
3 11 0.04
4 11 0.04
5 11 0.00
6 11 0.04
7 11 0.01
8 11 0.00
9 11 0.03
10 11 0.04
11 11 0.67
12 11 0.04
13 11 0.03
14 11 0.00
15 11 0.01
16 11 0.00
0 12 0.03
1 12 0.01
2 12 0.01
3 12 0.04
4 12 0.01
5 12 0.00
6 12 0.00
7 12 0.00
8 12 0.00
9 12 0.03
10 12 0.01
11 12 0.01
12 12 0.81
13 12 0.00
14 12 0.01
15 12 0.04
16 12 0.00
0 13 0.01
1 13 0.04
2 13 0.01
3 13 0.02
4 13 0.01
5 13 0.00
6 13 0.12
7 13 0.01
8 13 0.09
9 13 0.02
10 13 0.03
11 13 0.01
12 13 0.00
13 13 0.57
14 13 0.03
15 13 0.01
16 13 0.04
0 14 0.02
1 14 0.01
2 14 0.01
3 14 0.02
4 14 0.01
5 14 0.00
6 14 0.00
7 14 0.04
8 14 0.01
9 14 0.01
10 14 0.00
11 14 0.00
12 14 0.01
13 14 0.06
14 14 0.78
15 14 0.00
16 14 0.03
0 15 0.09
1 15 0.04
2 15 0.01
3 15 0.03
4 15 0.00
5 15 0.01
6 15 0.01
7 15 0.00
8 15 0.01
9 15 0.11
10 15 0.05
11 15 0.00
12 15 0.04
13 15 0.00
14 15 0.01
15 15 0.61
16 15 0.00
0 16 0.01
1 16 0.01
2 16 0.04
3 16 0.00
4 16 0.00
5 16 0.00
6 16 0.01
7 16 0.00
8 16 0.23
9 16 0.01
10 16 0.00
11 16 0.00
12 16 0.00
13 16 0.06
14 16 0.01
15 16 0.01
16 16 0.63
};
\end{axis}
%\draw[black,thick] (2.625,5.8) circle(0.5);
%\draw[red,dashed,thick] (3.675,5.8) circle(0.5);
\end{tikzpicture}
\caption{Polar KNN HOG confusion matrix (accuracy~0.7009)}\label{fig:conf_KHP}
\end{figure}
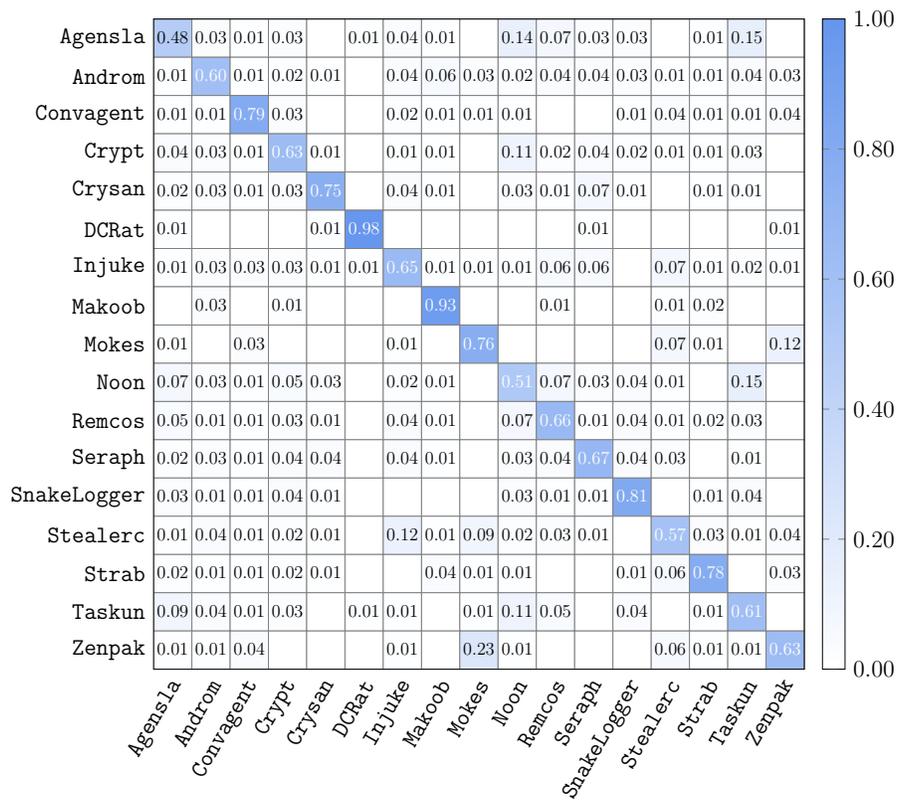

\end{document}